\documentclass[letterpaper]{article}
\usepackage[latin9]{inputenc}
\usepackage{prettyref}
\usepackage{mathtools}
\usepackage{amsmath}
\usepackage{amsthm}
\usepackage{amssymb}
\usepackage{graphicx}
\usepackage{microtype}
\usepackage[unicode=true,
 bookmarks=false,
 breaklinks=false,pdfborder={0 0 1},backref=section,colorlinks=false]
 {hyperref}

\makeatletter

\pdfpageheight\paperheight
\pdfpagewidth\paperwidth

\newcommand{\lyxdot}{.}

\usepackage{subfigure}

\usepackage{empheq}

\usepackage[accepted]{./icml2024}

\usepackage{amsthm}
\usepackage{cuted}

\usepackage[capitalize,noabbrev]{cleveref}

\theoremstyle{plain}

\theoremstyle{definition}
\theoremstyle{remark}

\AtBeginDocument{%
\newrefformat{fig}{\hyperref[#1]{Fig.~\ref{#1}}}
}

\makeatother

\begin{document}
\global\long\def\tr{\mathrm{tr}}%
\global\long\def\T{\mathsf{T}}%
\global\long\def\N{\mathcal{N}}%
\global\long\def\W{\mathcal{W}}%
\global\long\def\D{\mathcal{D}}%
\global\long\def\tC{\tilde{C}}%
\global\long\def\const{\mathrm{const}.}%
\global\long\def\I{\mathbb{I}}%
\global\long\def\order{\mathcal{O}}%
\global\long\def\bX{\mathbf{X}}%
\global\long\def\bY{\mathbf{Y}}%
\global\long\def\S{\mathcal{S}}%
\global\long\def\cP{\mathcal{P}}%
\global\long\def\cL{\mathcal{L}}%
\global\long\def\bR{\mathbb{R}}%
\icmltitlerunning{Critical feature learning in deep neural networks} 

\twocolumn[
\icmltitle{Critical feature learning in deep neural networks}

\icmlsetsymbol{equal}{*}

\begin{icmlauthorlist}
\icmlauthor{Kirsten Fischer}{equal,fzj,rwth_phd}
\icmlauthor{Javed Lindner}{equal,fzj,rwth,rwth_cosmo}
\icmlauthor{David Dahmen}{fzj}
\icmlauthor{Zohar Ringel}{huji}
\icmlauthor{Michael Kr\"amer}{rwth_cosmo}
\icmlauthor{Moritz Helias}{fzj,rwth}
\end{icmlauthorlist}

\icmlaffiliation{fzj}{
Institute for Advanced Simulation (IAS-6), Computational and Systems Neuroscience, Jülich Research Centre, Jülich, Germany}
\icmlaffiliation{rwth}{Department of Physics, RWTH Aachen University, Aachen, Germany}
\icmlaffiliation{rwth_phd}{RWTH Aachen University, Aachen, Germany}
\icmlaffiliation{rwth_cosmo}{Institute for Theoretical Particle Physics and Cosmology, RWTH Aachen University, Aachen, Germany}
\icmlaffiliation{huji}{The Racah Institute of Physics, The Hebrew University of Jerusalem, Jerusalem, Israel}

\icmlcorrespondingauthor{Kirsten Fischer}{ki.fischer@fz-juelich.de}
\icmlcorrespondingauthor{Javed Linder}{javed.lindner@rwth-aachen.de}

\vskip 0.3in
]
\printAffiliationsAndNotice{\icmlEqualContribution}
\begin{abstract}
A key property of neural networks driving their success is their
ability to learn features from data. Understanding feature learning
from a theoretical viewpoint is an emerging field with many open questions.
In this work we capture finite-width effects with a systematic theory
of network kernels in deep non-linear neural networks. We show that
the Bayesian prior of the network can be written in closed form as
a superposition of Gaussian processes, whose kernels are distributed
with a variance that depends inversely on the network width $N$.
A large deviation approach, which is exact in the proportional limit
for the number of data points $P=\alpha N\to\infty$, yields a pair
of forward-backward equations for the maximum a posteriori kernels
in all layers at once. We study their solutions perturbatively to
demonstrate how the backward propagation across layers aligns kernels
with the target. An alternative field-theoretic formulation shows
that kernel adaptation of the Bayesian posterior at finite-width results
from fluctuations in the prior: larger fluctuations correspond to
a more flexible network prior and thus enable stronger adaptation
to data. We thus find a bridge between the classical edge-of-chaos
NNGP theory and feature learning, exposing an intricate interplay
between criticality, response functions, and feature scale.
\end{abstract}

\section{Introduction}

A central quest of the theory of deep learning is to understand
the inductive bias of network architectures, which is their ability
to find solutions that generalize well despite networks being highly
overparametrized. The regime of lazy learning \cite{Chizat19_neurips},
in which the network width $N\to\infty$ tends to infinity while the
number of training data points $P$ stays constant, is well understood
in terms of the neural network Gaussian process (NNGP) \cite{Lee18}
and the neural tangent kernel (NTK) \cite{Jacot18_8580}. The NNGP
is, however, identical to training the readout weights only \cite{Lee19_neurips,Yang19}.
The NNGP kernel follows from the central limit theorem applied to
random networks, neglecting any adaptation to the data. While the
NTK describes the evolution of weights in all layers, it applies to
the case of small learning rates, effectively linearizing the mapping
between weights and outputs around the point of initialization. Consequently,
weights change only negligibly compared to initialization.

At finite network width or when keeping the ratio $\alpha=P/N$ constant
and taking the limit $N\to\infty$, the intermediate network layers
adapt to data; they learn ``features''. Feature learning typically
outperforms networks in the lazy regime \cite{Novak19_iclr,Lee20_ad086f59,Geiger20_113301,Petrini22_arxiv}
and is also required to understand transfer learning, the central
mechanism that enables modern foundation models \cite{Bommasani22_arxiv}.

\begin{figure}[tb]
\vspace{0.2in}
\includegraphics[width=1\columnwidth]{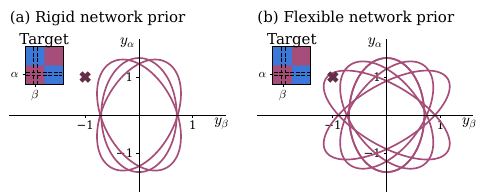}\caption{Larger kernel fluctuations enable stronger feature learning. The network
prior is a superposition of Gaussians given by $f\sim\int\,\protect\N(0,C)\,p(C)\,dC$
(pink ellipses). Depending on the network hyperparameters, the distribution
of kernels is more concentrated (a) or wider (b), corresponding to
smaller or larger kernel fluctuations. The target kernel is given
by $YY^{\mathsf{T}}$ (inset); the target value for the indicated
example samples $\alpha,\,\beta$ (dashed lines in inset) from different
classes lies at $(+1,-1)$ (red cross). In the Bayesian posterior
Gaussian components are reweighed according to the data. Larger fluctuations
in (b) allow stronger adaptation to data, leading to richer feature
learning.\label{fig:Visual-abstract}}
\vspace{0.2in}
\end{figure}

We here derive a theory of data-adaptive kernels in deep non-linear
networks trained in a Bayesian manner. We show that the prior for
the network outputs $f$ can be written as a superposition of Gaussian
processes $f\sim\int\,\N(0,C)\,p(C)\,dC$. Feature learning may be
understood as a reweighing of different components $\N(0,C)$ within
this prior ensemble according to the evidence $p(Y|C)=\N(Y|0,C)$
of the training labels $Y$. As a result, the posterior is dominated
by those Gaussian components $\N(0,C)$ that have a high evidence.
A wide distribution $p(C)$ leads to a rich prior (see \prettyref{fig:Visual-abstract})
and thereby enables strong adaptation to the training data. This view
allows us to connect feature learning to the notion of criticality:
these are points in hyperparameter space where the distribution $p(C)$
becomes particularly wide because the network is at the verge of transitioning
between two qualitatively different regimes.

The main contributions of this work are:
\begin{itemize}
\item an exact decomposition of the network prior into a superposition of
Gaussian processes, whose covariances are distributed with width of
$\order(N^{-1})$;
\item exact expressions for the Bayesian maximum a posteriori kernels in
the proportional limit $N,\,P\to\infty$ with $P/N=\alpha$ that follow
from a large deviation approach, yielding a set of forward-backward
self-consistent kernel propagation equations;
\item demonstration that a perturbative evaluation of the forward-backward
propagation of kernels captures feature learning in trained networks;
\item the discovery of a tight link between fluctuations near a critical
point and the ability of the network to show feature learning, uncovering
the driving mechanism behind feature learning as a tradeoff between
criticality and feature learning scale of the network output.
\end{itemize}

\section{Related works}

Previous work has investigated deep networks within the Gaussian process
limit for infinite width $N\to\infty$ \cite{Schoenholz17_iclr,Lee18}.
\cite{Schoenholz17_iclr} found optimal backpropagation of signals
and gradients when initializing networks at the critical point, the
transition to chaos \cite{molgedey92_3717}. Our work goes beyond
the Gaussian process limit by studying the joint limit $N\to\infty$,
$P\to\infty$ with $P/N=\alpha$ fixed. This limit has been investigated
with tools from statistical mechanics in deep linear networks \cite{Li21_031059},
where kernels adapt to data by only changing their overall scale compared
to the NNGP limit. A rigorous non-asymptotic solution for deep linear
networks in terms of Meijer-G functions \cite{Hanin23} has shown
that the posterior of infinitely deep linear networks with data-agnostic
priors is the same as that of shallow networks with evidence-maximizing
data-dependent priors. For a teacher-student setting, \cite{ZavatoneVeth22_064118}
show that in deep linear networks feature learning corrections to
the generalization error result from perturbation corrections only
at quadratic order or higher. For deep kernel machines, \cite{Yang23_39380}
find a similar trade-off between network prior and data term as we
do; in contrast to our work they study a different limit with $P$
fixed and train on $N$ copies of the data. Their main results can
be obtained from ours in the special case of deep linear networks
(see \prettyref{app:linear_net}); most importantly for non-linear
networks they require the use of normalizing flows to capture non-Gaussian
effects while our work provides a mechanistic understanding of such
effects.

Previous theoretical work on non-linear networks of finite width $N<\infty$
has employed three different approximation techniques. First, a perturbative
approach that computes corrections where the non-linear terms constitute
the expansion parameter \cite{Halverson21_035002}. Second, a perturbative
approach based on the Edgeworth expansion that uses the strength of
the non-Gaussian cumulants as an expansion parameter. These corrections
are computed either in the framework of gradient-based training \cite{Dyer20_ICLR,Huang20_4542,Aitken20_06687,Roberts22,Bordelon23_114009}
or Bayesian inference \cite{Yaida20,Antognini19_arxiv,Naveh21_064301,Cohen21_023034,Roberts22}.
\cite{ZavatoneVeth21_NeurIPS_II} derive a general form of finite-width
corrections, resulting from the linear readout layer and the quadratic
loss function. Third, non-perturbative Bayesian approaches \cite{Naveh21_NeurIPS,Seroussi23_908,Pacelli23_1497,Cui23_6468},
that derive self-consistency equations either by saddle-point integration
or by variational methods to obtain the Bayesian posterior. \cite{Cui23_6468}
exploits the Nishimori conditions that hold for Bayes-optimal inference,
where student and teacher have the same architecture and the student
uses the teacher's weight distribution as a prior; the latter is assumed
Gaussian i.i.d., which allows them to use the Gaussian equivalence
principle \cite{Goldt20_14709} to obtain closed-form solutions. Our
work is most closely related to these non-perturbative Bayesian approaches.
The qualitative difference is that we describe the trade-off between
the data term and the network prior in a large deviation approach
that is exact in the proportional limit and we do not require particular
assumptions on the data statistics. Our alternative field-theoretical
view connects this approach to finite-size fluctuations, by which
we discover a link between feature learning corrections and criticality
in deep networks.

\section{Feature learning theory of Bayesian network posterior\label{sec:theory}}

We consider a fully-connected, deep, feed-forward network
\begin{align}
h_{\alpha}^{(0)} & =W^{(0)}x_{\alpha}+b^{(0)}\,,\nonumber \\
h_{\alpha}^{(l)} & =W^{(l)}\phi\left(h_{\alpha}^{(l-1)}\right)+b^{(l)}\quad l=1,\ldots,L,\label{eq:SecTheory_NetworkArchitecture}\\
f_{\alpha} & =h_{\alpha}^{(L)},\nonumber 
\end{align}
with data indices $\alpha\in\{1,\dots,P\}$, where $P$ denotes the
number of training samples. We have inputs $x_{\alpha}\in\mathbb{R}^{D}$,
hidden states $h_{\alpha}^{(l)}\in\mathbb{R}^{N}$, and network output
$f_{\alpha}\in\mathbb{R}$. To ease notation, we assume identical
width $N$ for all layers. We derive the theoretical framework for
arbitrary activation functions $\phi:\mathbb{R\mapsto\mathbb{R}}$,
but consider $\phi(x)=\mathrm{erf}(x)$ for quantitative results in
subsequent sections. Further we assume Gaussian i.i.d. priors for
all weights ${W^{(0)}\in\mathbb{R}^{N\times D},\,W^{(l)}\in\mathbb{R}^{N\times N},\,W^{(L)}\in\mathbb{R}^{1\times N}}$
and biases ${b^{(l)}\in\mathbb{R}^{N},\,b^{(L)}\in\mathbb{R}}$ so
that ${W_{ij}^{(0)}\stackrel{\text{i.i.d.}}{\sim}\mathcal{N}\left(0,g_{0}/D\right),\,W_{ij}^{(l)}\stackrel{\text{i.i.d.}}{\sim}\mathcal{N}\left(0,g_{l}/N\right)}$
for $i,\,j=1,\dots,N$ and $l=1,\dots,L-1$, ${W_{i}^{(L)}\overset{\mathrm{\text{i.i.d.}}}{\sim}\mathcal{N}\left(0,g_{L}/N\right)}$
and ${b_{i}^{(l)}\stackrel{\text{i.i.d.}}{\sim}\mathcal{N}(0,g_{b})}$
for ${i=1,\dots,N}$ and ${l=0,\dots,L}$. We study the Bayesian posterior
distribution conditioned on a training data set consisting of inputs
${X=(x_{\alpha})_{\alpha=1,\ldots,P}}$ and corresponding labels ${Y=(y_{\alpha})_{\alpha=1,\ldots,P}}$
as in \cite{Naveh20_01190,Li21_031059,Segadlo22_103401}. This can
alternatively be seen as training the network with stochastic Langevin
dynamics (see Appendix \prettyref{app:langevin}).

\subsection{Network prior as superposition of Gaussians}

Assuming sample-wise i.i.d. Gaussian regularization noise of variance
$\kappa$, the network prior ${p(Y\vert X)=\int\,\prod_{\alpha=1}^{P}\N(y_{\alpha}|f_{\alpha},\kappa)\,p(f|X)\,df}$
with network outputs $f=(f_{\alpha})_{\alpha=1,\dots,P}$ follows
from the network mapping \prettyref{eq:SecTheory_NetworkArchitecture}
by enforcing the network architecture through Dirac distributions,
taking the expectation over all parameters $\Theta=\{W^{(l)},b^{(l)}\}_{l}$,
and introducing auxiliary variables $C_{\alpha\beta}^{(l)}\coloneqq g_{l}/N\,\phi_{\alpha}^{(l-1)}\cdot\phi_{\beta}^{(l-1)\T}+g_{b}$
with the shorthand $\phi_{\alpha i}^{(l)}=\phi\left(h_{\alpha i}^{(l)}\right)$,
similar to \cite{Segadlo22_103401} (see Appendix \prettyref{app:theory})
\begin{align}
p(Y|X) & =\int\D C\,\mathcal{N}\left(Y\vert0,C^{(L)}+\kappa\I\right)\,p(C),\label{eq:network_prior}\\
p(C) & =\int\D\tC\,\exp\left(-\tr\,\tC^{\T}C+\W(\tC|C)\right),\label{eq:p_C}
\end{align}
where $\tilde{C}^{(l)}$ is the conjugate kernel to $C^{(l)}$ and
${\tr\,\tC^{\T}C=\sum_{\alpha\beta l}\tC_{\alpha\beta}^{(l)}C_{\alpha\beta}^{(l)}}$.
This expression shows that the network output is a superposition of
centered Gaussian processes $\mathcal{N}\left(0,C^{(L)}+\kappa\I\right)$.
Its covariance depends on $C^{(L)}$ that itself is distributed as
${p(C^{(L)})=\int dC^{(1\le l<L)}\,p(C)}$, where the joint distribution
${p(C)=p(C^{(L)}|C^{(L-1)})\cdots p(C^{(1)}|C^{(0)})}$ of all $C^{(1\le l\le L)}$
decomposes into a chain of conditionals. The distribution $p(C^{(L)})$
is given by its cumulant generating function
\begin{align}
 & \W(\tC|C)\label{eq:cum_W}\\
 & =N\,\sum_{l=0}^{L-1}\ln\,\Big\langle\exp\big(\frac{g_{l+1}}{N}\phi^{(l)\T}\tC^{(l+1)}\phi^{(l)}\big)\Big\rangle_{\N(0,C^{(l)})}\nonumber \\
 & \quad+\tC g_{b}+\tC^{(0)\T}C^{(0)},\nonumber 
\end{align}
where $\phi^{\T}\tC\phi=\sum_{\alpha\beta}\phi_{\alpha}\tC_{\alpha\beta}\phi_{\beta}$.
We write here and in the following $\langle\ldots\rangle_{\N(0,C^{(l)})}\equiv\langle\ldots\rangle_{h^{(l)}\sim\N(0,C^{(l)})}$
for the Gaussian expectation value of the activations $h^{(l)}$ with
regard to a centered Gaussian measure with covariance matrix $C^{(l)}\in\mathbb{R}^{P\times P}$
and denote as
\begin{align}
C^{(0)} & =\frac{g_{0}}{D}\,XX^{\T}+g_{b}\label{eq:C_0}
\end{align}
the Gaussian kernel after the readin layer. The network prior \eqref{eq:network_prior},
written as a superposition of Gaussians, is an exact result. We next
determine the maximum a posteriori (MAP) estimate for the $C^{(l)}$.

\subsection{Large deviation approach for the maximum a posteriori kernel}

The cumulant-generating function \eqref{eq:cum_W} has what is known
as a scaling form \cite{Touchette09} $\W(\tC)=N\,\lambda(\tC/N)$
with an $N$-independent function $\lambda$; thus its $k$-th cumulant
scales with $1/N^{k-1}$ so that $C$ concentrates as $N\to\infty$
around its mean. So while the kernel of the input layer $C^{(0)}$
is deterministic, all subsequent auxiliary variables $C^{(1\leq l\le L)}$
are fluctuating quantities with a variance of order ${\W^{\prime\prime}\sim\order(N^{-1})}$.
The scaling form at large $N$ implies that we may approximate the
integral over $\tC^{(l)}$ in \eqref{eq:p_C} for $1\le l\le L$ by
the G\"artner-Ellis theorem to obtain a large deviation principle
(l.d.p.)
\begin{align}
 & -\ln p(C^{(l+1)}|C^{(l)})\label{eq:rate_function-1}\\
 & \stackrel{\text{l.d.p.}}{\simeq}\sup_{\tC^{(l+1)}}\,\tr\,\tC^{(l+1)\T}C^{(l+1)}-\W(\tC^{(l+1)}|C^{(l)})\nonumber \\
 & \eqqcolon\Gamma(C^{(l+1)}|C^{(l)}),\nonumber 
\end{align}
expressed in terms of the rate function $\Gamma$ \cite{Touchette09}.
To provide more intuition for the rate functions $\Gamma$, we show
in Appendix \prettyref{app:linear_net} that for linear networks the
rate function reduces to the Kullback-Leibler divergence between the
Gaussian distributions of the two adjacent layers' activations. Thus,
the prior has the tendency to keep the distributions in adjacent layers
close to one another. The joint probability $p(C)$ in \eqref{eq:p_C}
then decomposes as
\begin{align*}
\ln p(C) & =\ln p(C^{(L)}|C^{(L-1)})\cdots p(C^{(1)}|C^{(0)})\\
 & \stackrel{\text{l.d.p.}}{\simeq}-\sum_{l=1}^{L}\Gamma(C^{(l)}|C^{(l-1)})=:-\Gamma(C).
\end{align*}
The supremum condition in \eqref{eq:rate_function-1} amounts to
\begin{align}
C_{\alpha\beta}^{(l+1)}\equiv & \frac{\partial\W}{\partial\tilde{C}_{\alpha\beta}^{(l+1)}}=g_{l+1}\,\big\langle\phi_{\alpha}^{(l)}\phi_{\beta}^{(l)}\big\rangle{}_{\mathcal{P}^{(l)}}+g_{b},\label{eq:prop_general}\\
\big\langle\ldots\big\rangle_{\mathcal{P}^{(l)}} & \propto\left\langle \ldots\,\exp\left(\frac{g_{l}}{N}\phi^{(l)\T}\tilde{C}^{(l+1)}\phi^{(l)}\right)\right\rangle _{\N(0,C^{(l)})},\label{eq:measure_non_Gauss-1}
\end{align}
where we defined the non-Gaussian measure $\langle\ldots\rangle_{\mathcal{P}^{(l)}}\equiv\langle\ldots\rangle_{h^{(l)}\sim\mathcal{P}(\tC^{(l+1)},C^{(l)})}$
and its proportionality constant is given by the proper normalization
(for details see \prettyref{app:Maximum-network-kernels}).

We condition on the training data, enforcing the labels $\{y_{\alpha}\}_{\alpha}$,
to obtain the posterior distribution for $C$ as ${p(C|Y)\propto p(Y,C)\equiv\mathcal{N}\left(Y\vert0,C^{(L)}+\kappa\I\right)\,p(C)}$,
where we read off the latter form of the joint density of $Y$ and
$C$ from \eqref{eq:network_prior}. We are interested in the maximum
a posteriori estimate for $C$, which is given by the stationary points
of
\begin{align}
\S(C) & \coloneqq\ln p(C|Y)\stackrel{\text{l.d.p.}}{\simeq}\mathcal{S}_{\mathrm{D}}(C^{(L)})-\Gamma(C)+\circ,\label{eq:action_C}\\
\mathcal{S}_{\mathrm{D}}(C^{(L)}) & \coloneqq-\frac{1}{2}Y^{\T}(C^{(L)}+\kappa\I)^{-1}Y\nonumber \\
 & \phantom{\coloneqq}-\frac{1}{2}\ln\det(C^{(L)}+\kappa\I),\nonumber 
\end{align}
where we dropped terms $\circ$ that are independent of $C$ and approximated
$p(C)$ by its rate function \eqref{eq:rate_function-1}. The exponent
$\S(C)$ has two terms: The log likelihood of the training labels
$\mathcal{S}_{\mathrm{D}}(C^{(L)})\sim\order(P)$ and the rate function
$-\Gamma(C)$ which arises from the network prior. It is easy to see
from \eqref{eq:cum_W} that its Legendre transform $\Gamma$ scales
with $\order(N)$.

The stationary point $\partial S(C)/\partial C^{(L)}\stackrel{!}{=}0$
of \eqref{eq:action_C} with regard to $C^{(L)}$ therefore arises
from a trade-off between the network prior term in the form of $\Gamma$
and the data term $\mathcal{S}_{\mathrm{D}}$. In the last layer this
yields
\begin{align}
\tilde{C}^{(L)} & =\frac{1}{2}(C^{(L)}+\kappa\I)^{-1}YY^{\T}(C^{(L)}+\kappa\I)^{-1}\label{eq:C_tilde_final}\\
 & \phantom{=}-\frac{1}{2}(C^{(L)}+\kappa\I)^{-1},\nonumber 
\end{align}
which expresses the value of $\tC^{(L)}$ in the final layer in terms
of the value of $C^{(L)}$ and the training labels $Y$. We further
show in Appendix \prettyref{app:conj_kernel_discrepancy} that the
conjugate kernel $\tilde{C}^{(L)}$ can be expressed in terms of the
second moment of the discrepancies between target and the network
output and its trace measures the training loss. Using Price's theorem
(see Appendix \prettyref{app:Generalization-of-Price}) and the fundamental
property of the Legendre transform in \eqref{eq:rate_function-1},
stationarity $\partial S(C)/\partial C^{(l)}\stackrel{!}{=}0$ yields
for intermediate network layers $1\le l<L$
\begin{align}
\tilde{C}_{\alpha\beta}^{(l)} & =-\frac{\partial\Gamma(C^{(l+1)}|C^{(l)})}{\partial C_{\alpha\beta}^{(l)}}\stackrel{\text{Legendre}}{\equiv}\frac{\partial\W(\tC^{(l+1)}|C^{(l)})}{\partial C_{\alpha\beta}^{(l)}}\nonumber \\
 & \stackrel{\text{Price's theorem}}{=}g_{l+1}\,\tilde{C}_{\alpha\beta}^{(l+1)}\,\Big\langle\left(\phi_{\alpha}^{(l)}\right)^{\prime}\left(\phi_{\beta}^{(l)}\right)^{\prime}\Big\rangle_{\mathcal{P}^{(l)}}\label{eq:CFL_Full_Tilde_Kernel_Propagation}\\
 & +\delta_{\alpha\beta}\,g_{l+1}\,\tilde{C}_{\alpha\alpha}^{(l+1)}\,\Big\langle\left(\phi_{\alpha}^{(l)}\right)^{\prime\prime}\phi_{\alpha}^{(l)}\Big\rangle_{\mathcal{P}^{(l)}}+\order(N^{-1}),\nonumber 
\end{align}
where we do not spell out terms $\propto\order(N^{-1})$ (the form
of which is given in Appendix \prettyref{app:theory}). This equation
thus gives $\tC^{(l)}$ in terms of $\tC^{(l+1)}$ and $C^{(l)}$.
The conjugate kernels $\tilde{C}^{(l)}$ propagate information about
the relation between inputs and outputs backwards across layers. By
\eqref{eq:C_tilde_final}, these are driven by the difference of two
terms: The conjugate kernel of the output layer $\tilde{C}^{(L)}$
measures the mismatch between output kernel $C^{(L)}$ and target
kernel $YY^{\mathsf{T}}$ and can be interpreted as an error signal.
In the following we will see that this error signal on the level of
the kernel is backpropagated by the backward response function and
exhibits an exponential decay over layers (similar to the response
studied in \cite{Schoenholz17_iclr}), indicating how information
backpropagates within the network.

\subsection{Forward-backward kernel propagation in the proportional limit}

The main result of the previous section is the pair of equations \eqref{eq:prop_general}
and \eqref{eq:CFL_Full_Tilde_Kernel_Propagation}
\begin{align}
C^{(l+1)} & \stackrel{(\ref{eq:prop_general})}{=}F(C^{(l)},\tC^{(l+1)}),\label{eq:prop_forward_ff_b}\\
\tilde{C}^{(l)} & \stackrel{(\ref{eq:CFL_Full_Tilde_Kernel_Propagation})}{=}G(C^{(l)},\tC^{(l+1)})\,\tC^{(l+1)},\label{eq:backward_tilde}
\end{align}
with initial and final conditions, respectively, given by \eqref{eq:C_0}
and \eqref{eq:C_tilde_final}, rewritten as
\begin{align}
\tC^{(L)} & =\frac{1}{2}(C^{(L)}+\kappa\I)^{-1}(YY^{\T}-C^{(L)}-\kappa\I)(C^{(L)}+\kappa\I)^{-1}.\label{eq:tilde_C_final_rewritten}
\end{align}
This set of equations (including the term $\order(N^{-1})$ in \eqref{eq:CFL_Full_Tilde_Kernel_Propagation})
is exact in the proportional limit $P=\alpha N\to\infty$; this is
so because the rate function \eqref{eq:rate_function-1} approximates
$-\ln p(C)$ correct up to additive constants, so that the stationary
points correctly determine the mode of the posterior for $C^{(l)}$.

The first equation \eqref{eq:prop_forward_ff_b} maps the MAP kernel
$C^{(l)}\mapsto C^{(l+1)}$ forward through the network. This mapping
in the $l$-th layer depends on $\tC^{(l+1)}$. This result is similar
to the NNGP limit \cite{Neal96,Williams96_ae5e3ce4,Lee17_00165}:
We in fact recover the latter in the case of a fixed number of training
samples $P$ and an infinitely wide network with $N\rightarrow\infty$
from the stationary point of \eqref{eq:action_C} which is then approximated
as $\mathcal{S}(C)\simeq-\Gamma(C)$. In this limit it follows from
the equation of state $\partial\Gamma(C)/\partial C_{\alpha\beta}^{(l)}=\tC_{\alpha\beta}^{(l)}$
that $\tC\equiv0$ vanishes and the measure \eqref{eq:measure_non_Gauss-1}
becomes the Gaussian measure with covariance $C^{(l)}$. In consequence
\eqref{eq:prop_general} reduces to the NNGP $C_{\alpha\beta}^{(l+1)}=g_{l+1}\left\langle \phi_{\alpha}^{(l)}\phi_{\beta}^{(l)}\right\rangle _{\N(0,C^{(l)})}+g_{b}$.
Among others, \cite{Yang20_14522} show that the NNGP limit fails
to capture feature learning which appears in neural networks in the
rich regime. Furthermore, we show in Appendix \prettyref{app:ntk}
that the NTK is contained in our framework as a special case that
assumes a linear dependence of the output on all layer's weights.

The here presented theoretical framework captures feature learning
in settings where the log-likelihood of the data $S_{D}$ is not negligible
compared to $\Gamma$ in \eqref{eq:action_C}; so either in the limit
$N\to\infty$ when the number of data samples scales linearly $P=\alpha\,N$,
or when $N,P$ are both large but finite. In the latter case, feature
learning results from the leading-order fluctuation corrections in
$N^{-1}$, as we show in the Appendix \prettyref{app:AppendixFluctuations}.
In both cases, the maximum a posteriori for $C$ balances the maximization
of the likelihood of the data $S_{D}$ and the maximization of the
log probability $-\Gamma(C)$ from the prior, leading to the equation
\eqref{eq:backward_tilde}, which propagates $\tC^{(l+1)}\to\tC^{(l)}$
backwards and in addition depends on $C^{(l)}$. In particular, we
will see that the data term in \eqref{eq:action_C} leads to the correction
of the output kernel $C^{(L)}$ towards the target kernel $YY^{\mathsf{T}}$
in \eqref{eq:C_tilde_final} and \eqref{eq:backward_tilde}. Such
an alignment means that the output of the network more closely reproduces
the outputs given by the training data. Such a term is absent both
in the NNGP and the NTK, both of which only depend on the training
data inputs $x$, and are hence unable to form relationships for the
input-label pairs $(x,y)$. In Appendix \prettyref{app:target_adaptation},
we show for deep linear networks that to leading order the correction
terms add a rank one contribution $YY^{\mathsf{T}}$ to the kernel.

Instead of taking expectations over the standard Gaussian measure
with covariance $C^{(l)}$ as in the NNGP, the forward propagation
\eqref{eq:prop_forward_ff_b} here employs a non-Gaussian probability
measure $\eqref{eq:measure_non_Gauss-1}$ that involves the activation
function $\phi$, the kernels $C^{(l)}$, and the conjugate kernel
$\tilde{C}^{(l+1)}$.

Finally, the value for $\tilde{C}^{(L)}$ given by \eqref{eq:tilde_C_final_rewritten}
allows for an intuitive interpretation: $C^{(L)}+\kappa\I=YY^{\T}$
implies $\tilde{C}^{(L)}=0$ and, subsequently by the linear dependence
on $\tC^{(l+1)}$ in \eqref{eq:backward_tilde}, that all vanish,
$\tilde{C}^{(1\le l\le L)}=0$. Hence at this point $\tC$ does not
drive further adaptation towards the target, as the output kernel
is already perfectly aligned to the desired target.

\subsection{Perturbative, leading-order solution of the forward-backward equations}

The presented approach does not depend on the choice of the activation
function $\phi$. For general activation functions $\phi$, however,
the exact expressions for the feature learning limit are hardly tractable
due to the non-Gaussian expectation value with regard to the measure
$\eqref{eq:measure_non_Gauss-1}$. The non-Gaussianity in the measure
$\eqref{eq:measure_non_Gauss-1}$ comes in the form of $\frac{g_{l}}{N}\phi_{\alpha}^{(l)}\tC_{\alpha\beta}^{(l+1)}\phi_{\beta}^{(l)}$
in the exponent, so the magnitude of the entries are diminished by
$N^{-1}$ compared to those of $-\frac{1}{2}h_{\alpha}^{(l)}[C^{(l)}]_{\alpha\beta}h_{\alpha}^{(l)}$
from the Gaussian part of the measure.  So expanding in $N^{-1}$,
which amounts to expanding to linear order in $\tC$, we may replace
the forward propagation \eqref{eq:prop_general} by
\begin{align}
C_{\alpha\beta}^{(l+1)} & =g_{l+1}\,\Big\langle\phi_{\alpha}^{(l)}\phi_{\beta}^{(l)}\Big\rangle_{\N(0,C^{(l)})}+g_{b}\label{eq:perturbative_forward}\\
 & \quad+\frac{g_{l+1}^{2}}{N}\,\sum_{\gamma,\delta}V_{\alpha\beta,\gamma\delta}^{(l)}\,\tilde{C}_{\gamma\delta}^{(l+1)}+\mathcal{O}\left(N^{-2}\right),\nonumber \\
V_{\alpha\beta,\gamma\delta}^{(l)} & :=\big\langle\phi_{\alpha}^{(l)}\phi_{\beta}^{(l)}\phi_{\gamma}^{(l)}\phi_{\delta}^{(l)}\big\rangle_{\N(0,C^{(l)})}\label{eq:def_V}\\
 & -\big\langle\phi_{\alpha}^{(l)}\phi_{\beta}^{(l)}\big\rangle_{\N(0,C^{(l)})}\big\langle\phi_{\gamma}^{(l)}\phi_{\delta}^{(l)}\big\rangle_{\N(0,C^{(l)})},\nonumber 
\end{align}
where all expectation values are Gaussian $\langle\ldots\rangle_{\N(0,C^{(l)})}$.
Likewise, at the same order of approximation, we may replace in \eqref{eq:CFL_Full_Tilde_Kernel_Propagation}
$\langle\ldots\rangle_{\mathcal{P}^{(l)}}$ by $\langle\ldots\rangle_{\N(0,C^{(l)})}$,
because corrections come with at least one factor $N^{-1}$. While
the two-point integrals in \prettyref{eq:CFL_Full_Tilde_Kernel_Propagation},
\prettyref{eq:perturbative_forward} and \prettyref{eq:def_V} have
closed-form analytical solutions for certain non-linearities such
as $\phi=\mathrm{erf(}x)$, the four-point integral in \prettyref{eq:def_V}
is evaluated numerically (for details see Appendix \prettyref{app:details_implementation}).
The kernels $C^{(l+1)}$ thus receive a correction from the backpropagated
error signal in the form of $\tilde{C}^{(l+1)}$. The correction of
$C_{\alpha\beta}^{(l+1)}$ results not only from the kernel element
itself $\tilde{C}_{\alpha\beta}^{(l+1)}$, but also depends on its
interaction with all other data samples via the four-point interaction
term $\sum_{\gamma,\delta}V_{\alpha\beta,\gamma\delta}^{(l)}\,\tilde{C}_{\gamma\delta}^{(l+1)}$.

We solve the self-consistency equations for both kernels $C^{(l)}$
and conjugate kernels $\tilde{C}^{(l)}$ iteratively. Details on a
numerically stable implementation are given in Appendix \prettyref{app:details_implementation}.
All code is available under \cite{Fischer24_zenodo}.

\subsubsection*{Experiments}

\begin{figure*}[t]
\includegraphics{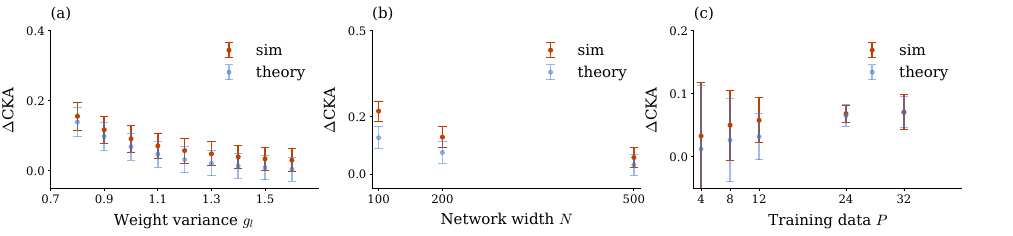}\vspace{0.2in}

\centering{}\caption{Comparison between theory and simulation for the XOR task. The difference
$\Delta\text{CKA}=\text{CKA}(C^{(l)},YY^{\mathsf{T}})-\text{CKA}(C^{\text{NNGP}},YY^{\mathsf{T}})$
measures kernel adaptation relative to the naive NNGP kernels. The
$\Delta\text{CKA}$ of the data-dependent kernels (blue: theory; red:
empirical) increases significantly (a) for smaller weight variance,
(b) for more narrow networks, and (c) for larger amounts of training
data. Parameters: XOR task with $\sigma^{2}=0.4,D=100,\,L=3$, (a)
$N=500,\,P=12$, (b) $P=12,\,g_{l}=1.2$, (c) $N=500,\,g_{l}=1.2$.
Results are averaged over $10$ training data sets and error bars
indicate standard deviation.\label{fig:cka_numerics_analytics_comparison}}
\vspace{0.2in}
\end{figure*}
We compare the obtained analytical results for the output kernel $C^{(L)}$
conditioned on the training data to the numerical implementation of
sampling the kernel $C_{\mathrm{\text{emp}}}^{(L)}$ from the posterior
distribution using Langevin stochastic gradient descent (see Appendix
\prettyref{app:langevin}). As a measure we use the centered kernel
alignments (CKA, see Appendix \prettyref{app:centred_kernel_alignment})
of both the analytical kernel $C^{(L)}$ and the Langevin sampled
kernels $C_{\mathrm{\text{emp}}}^{(L)}$ with the target kernel $YY^{\top}$
respectively. Since our framework does not presuppose any assumptions
on the data, we study two different tasks: XOR and binary classification
on MNIST digits; the numerical results match our theoretical expectations
consistently in both cases.

\paragraph*{XOR}

\cite{Refinetti21_8936} show that random feature models, which are
known to correspond to the NNGP \cite{Mei22}, are unable to solve
the non-linearly separable task XOR optimally. We study the XOR task
in a setting where neural networks exhibit feature learning compared
to random feature models \cite{Refinetti21_8936}. The feature-corrected
kernels that we obtain from our theory have a larger CKA than the
NNGP (see \prettyref{fig:cka_numerics_analytics_comparison}), indicating
that finite-width effects lead to kernel corrections in the direction
of the target kernel. Note that the CKA is by construction invariant
to a global rescaling of the kernel and instead captures the kernel
structure. Thus, the difference between NNGP and empirical kernels
is further numerical evidence that the kernels acquire structure beyond
a global rescaling, in contrast to deep linear networks \cite{Li21_031059}
and opposed to approximate results employing Gaussian equivalence
theory \cite{Pacelli23_1497,Baglioni24_arxiv}. We observe a stronger
kernel alignment for smaller weight variance $g_{l}$ (see \prettyref{fig:cka_numerics_analytics_comparison}(a)).
As expected, the feature-corrected kernels approach the NNGP limit
for $\alpha=P/N\rightarrow0$ when keeping $P$ fixed in \prettyref{fig:cka_numerics_analytics_comparison}(b).
Deviations for small $N$ in \prettyref{fig:cka_numerics_analytics_comparison}(b)
and in \prettyref{fig:cka_numerics_analytics_comparison}(c) for increasing
$P$ at fixed $N$ result from the perturbative treatment of $\tC$
in the numerical solution of the self-consistency equations, which
is strictly valid only for $\alpha=P/N\ll1$.

\paragraph*{MNIST}

\begin{figure}[tb]
\includegraphics[width=1\columnwidth]{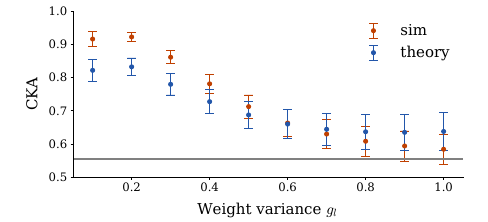}\vspace{0.2in}

\centering{}\caption{Comparison between theory and simulation for MNIST. The strength of
kernel adaptation measured as $\text{CKA}(C^{(l)},YY^{\mathsf{T}})$
shows a maximum that is consistent for theory (blue) and simulation
(red). Kernel adaptation increases significantly relative to the NNGP
(gray). Parameters: MNIST task with $L=2,\,N=2000$. Results are averaged
over $10$ training data sets and error bars indicate plus minus one
standard deviation.\label{fig:mnist}}
\vspace{0.2in}
\end{figure}

We study a binary classification task on MNIST \cite{Lecun1998} between
digits $0$ and $3$. The feature-corrected kernels obtained from
theory show increased kernel alignment with the target kernel $YY^{\mathsf{T}}$
compared to the NNGP, matching the behavior in neural networks trained
by Langevin dynamics (see \prettyref{fig:mnist}).

\section{Interplay between criticality and output scale\label{sec:criticality_learning}}

To understand the driving forces behind kernel adaptation to data
as presented in the previous section, we study the self-consistency
equations \prettyref{eq:CFL_Full_Tilde_Kernel_Propagation} for the
network kernels in detail. For the presented feature learning theory,
we reveal a link to fluctuations, the response function and the scales
within the network.

\subsection{Fluctuations lead to feature learning}

For the network prior in \prettyref{eq:network_prior}, we have defined
auxiliary variables $C_{\alpha\beta}^{(l)}=g_{l}/N\,\phi_{\alpha}^{(l-1)}\cdot\phi_{\beta}^{(l-1)\T}+g_{b}.$
For infinitely-wide networks $N\rightarrow\infty$ these quantities
concentrate, the scalar product over neuron indices becomes an expectation
value and we obtain the NNGP kernel given by $C_{\alpha\beta}^{(l),\mathrm{\text{NNGP}}}=g_{l}\,\langle\phi_{\alpha}^{(l-1)}\phi_{\beta}^{(l-1)}\rangle_{\mathcal{N}(0,C^{(l-1)})}+g_{b}$.
For large but finite network width $N<\infty$, the realizations of
the auxiliary variables measured from a particular network realization
$\Theta=\left\{ W^{(l)},\,b^{(l)}\right\} _{l}$ fluctuate around
the NNGP kernel
\[
C_{\alpha\beta}^{(l)}=C_{\alpha\beta}^{(l),\mathrm{\text{NNGP}}}+\delta C_{\alpha\beta}^{(l)}.
\]
We now show that corrections to the NNGP result derived from the perturbative
approach \eqref{eq:perturbative_forward} can alternatively be understood
as fluctuation corrections in a field-theoretic formulation (see Appendix
\prettyref{app:AppendixFluctuations}). We rewrite \prettyref{eq:network_prior}
and \prettyref{eq:p_C} as
\begin{align}
p(Y|X) & =\int\D C\int\D\tC\,\exp\big(\S(C,\tC)+\mathcal{S}_{\mathrm{D}}(C^{(L)}|Y)\big),\nonumber \\
\mathcal{S}_{\mathrm{D}}(C^{(L)}|Y) & =\ln\N(Y|0,C^{(L)}+\kappa\I).
\end{align}
We perform a Laplace approximation of $\exp\left(\S(C,\tC)\right)$
around its saddle point $C^{(l),\mathrm{*}}=C^{(l),\mathrm{\text{NNGP}}},\,\tilde{C}^{(l),*}=0$:
\begin{align}
 & p(Y|X)\label{eq:laplace_approx}\\
 & \simeq\int\D\delta C\,\int\D\delta\tC\,\exp\Big(\frac{1}{2}(\delta C,\delta\tC)^{\T}\mathcal{S}^{(2)}\,(\delta C,\delta\tC)\nonumber \\
 & \qquad\qquad\qquad\qquad\qquad\qquad+\mathcal{S}_{\mathrm{D}}(C_{\ast}^{(L)}+\delta C^{(L)}|Y)\Big),\nonumber 
\end{align}
where we write $\delta C=C-C^{*},\,\delta\tilde{C}=\tilde{C}-\tilde{C}^{*}$
and we denote the Hessian of $\S(C,\tC)$ as $S^{(2)}=\partial^{2}S/\partial(C,\tilde{C})$,
whose negative inverse yields the second cumulant of $(C,\tilde{C})$
\[
\left[-S^{(2)}\right]^{-1}=\left(\begin{array}{cc}
\langle\delta C\,\delta C\rangle & \ensuremath{\langle\delta C\,\delta\tC\rangle}\\
\ensuremath{\langle\delta\tC\,\delta C\rangle} & 0
\end{array}\right).
\]
Computing the saddle point of $\delta C$ in \eqref{eq:laplace_approx}
by taking the effect of $\partial\mathcal{S}_{\mathrm{D}}/\partial\delta C^{(L)}$
into account, we get
\begin{align}
\delta C_{\alpha\beta}^{(l)} & =g_{l}\,\sum_{\gamma\delta}\frac{\partial\langle\phi_{\alpha}^{(l-1)}\phi_{\beta}^{(l-1)}\rangle_{\mathcal{N}(0,C^{(l-1)})}}{\partial C_{\gamma\delta}^{(l-1)}}\,\delta C_{\gamma\delta}^{(l-1)}\label{eq:corrected_saddle}\\
 & \quad+g_{l}^{2}\,\sum_{\gamma\delta}V_{\alpha\beta,\gamma\delta}^{(l-1)}\,\delta\tilde{C}_{\gamma\delta}^{(l)}.\nonumber 
\end{align}
The first term in \eqref{eq:corrected_saddle} results from linearly
correcting the NNGP expression by the shift in $\delta C$
\begin{align*}
 & g_{l}\left\langle \phi_{\alpha}^{(l-1)}\phi_{\beta}^{(l-1)}\right\rangle _{\mathcal{N}(0,C^{(l-1)}+\delta C^{(l-1)})}\\
 & \quad=C_{\alpha\beta}^{(l),\mathrm{\text{NNGP}}}+\sum_{\gamma\delta}\frac{\partial C_{\alpha\beta}^{(l),\mathrm{\text{NNGP}}}}{\partial C_{\gamma\delta}^{(l-1)}}\delta C_{\gamma\delta}^{(l-1)}+\mathcal{O}(\delta C)^{2}.
\end{align*}
The second term in \eqref{eq:corrected_saddle} corresponds to the
corrections in \prettyref{eq:perturbative_forward}. By identifying
$g_{l}^{2}V_{\alpha\beta,\gamma\delta}^{(l-1)}$ given by \eqref{eq:def_V}
with the covariance of the auxiliary variables, we see how the feature
learning corrections in the self-consistency equations result from
fluctuation corrections. Thus, larger fluctuations lead to stronger
feature learning. Fluctuations become especially larger close to critical
points that mark phase transitions between qualitatively different
states in neural networks.

\subsection{Feature learning corrections close to criticality}

\begin{figure*}
\vspace{0.2in}

\includegraphics{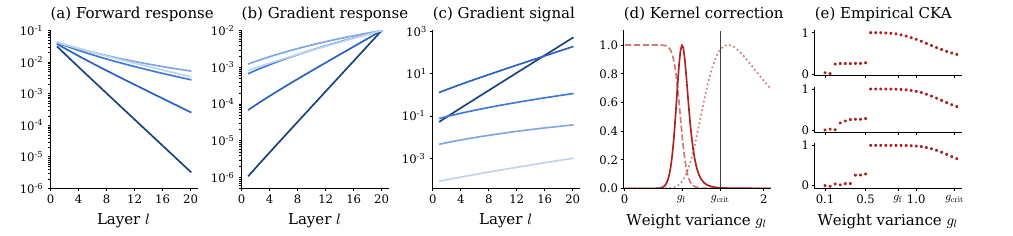}
\centering{}\caption{Finite-size effects close to criticality in feature learning. (a)-(b)
Forward response $\chi^{l,\rightarrow}$ and gradient response $\chi^{l,\leftarrow}$
measure relative signal propagation across network layers. The signal
propagates furthest close to criticality ($g_{l}$ increases from
dark to light blue). (c) Backpropagated conjugate kernel $\tilde{C}^{(l)}$
across network layers for varying weight variance $g_{l}$ (increasing
from dark to light). The kernel mismatch $\tilde{C}^{(L)}$ in the
output also depends on the weight variance $g_{l}$, so that the curves
of $\tilde{C}^{(l)}$ intersect at different depths. Larger $\tilde{C}^{(1)}$
in the first layer leads to stronger feature learning corrections
in \prettyref{eq:perturbative_forward}. (d) The kernel correction
term $\tilde{C}^{(0)}$ in the readin layer (solid line, slice for
$l=1$ in (c)) is composed of the gradient response $\chi^{1,\leftarrow}$
(dotted line, slice for $l=1$ in (b)) and the error signal $\tilde{C}^{(L)}$
(dashed line). Thus, strongest feature learning corrections occur
for a weight variance $g_{f}$ shifted away from the critical point
(vertical line) to smaller values. (e) CKA for trained networks between
$C_{\text{emp}}^{(l)}$ and $YY^{\mathsf{T}}$ ($l=10,\,15,\,20$
from top to bottom). Other parameters: XOR task with $\sigma^{2}=0.4$,
$g_{l}\in\left\{ 0.6,\,0.825,\,1.1,\,g_{\text{crit}}\approx1.38,\,2.2\right\} ,\,g_{b}=0.05,\,L=20,\,N=500\,,\kappa=10^{-3},\,P=12$.\label{fig:criticality_learning}}
\vspace{0.2in}
\end{figure*}
We study the relation of the self-consistency equations to criticality
in neural networks \cite{Schoenholz17_iclr}. The self-consistency
equations for the conjugate kernels \textbf{$\tilde{C}^{(l)}$} are
given by the iterative expression \prettyref{eq:CFL_Full_Tilde_Kernel_Propagation}.
Ultimately, the network kernels $C^{(l-1)}$ receive a correction
from the conjugate kernel \textbf{$\tilde{C}^{(l)}$}. For $\alpha\neq\beta$
it can be explicitly written to linear order as
\begin{align}
\tilde{C}_{\alpha\beta}^{(l)} & =\tilde{C}_{\alpha\beta}^{(L)}\,\chi_{\alpha\beta}^{(l),\leftarrow},\label{eq:backprop_C_tilde_total}\\
\chi_{\alpha\beta}^{(l),\leftarrow} & \coloneqq\prod_{s=l}^{L-1}g_{s+1}\,\Big\langle\left(\phi_{\alpha}^{(s)}\right)^{\prime}\left(\phi_{\beta}^{(s)}\right)^{\prime}\Big\rangle_{h^{(s)}\sim\mathcal{N}(0,C^{(s)})},\label{eq:backward_response}
\end{align}
where $\tC^{(L)}$ is given by \prettyref{eq:C_tilde_final}. As discussed
in the previous section, the term $\tilde{C}^{(L)}$ is related to
the mismatch between the output kernel $C^{(L)}$ and the target kernel
given by $YY^{\T}$. This error signal gets backpropagated from layer
to layer by the multiplicative terms in \prettyref{eq:backward_response}.
We identify $\chi^{(l),\leftarrow}$ as the gradient response function.
It is related to the forward response function \cite{Schoenholz17_iclr}
that measures how perturbations in the input kernel affect network
kernels in later layers $\chi_{\alpha\beta}^{l,\rightarrow}=\partial C_{\alpha\beta}^{(l)}/\partial C_{\alpha\beta}^{(0)}\big|_{C_{\text{NNGP}}}$.
Both response functions appear naturally in a field-theoretic description
of neural networks by considering Gaussian fluctuations of the kernels
as a first-order correction at finite width \cite{Segadlo22_103401,Fischer23_arxiv}.
The main difference between the forward and gradient response function
is that the signal perturbation leading to responses in the network
arises in different layers and thus propagates in opposite directions:
forward response propagates from input to output, gradient response
propagates from output to input.

For a particular set of network hyperparameters, both response functions
exhibit long-range correlation across layers (see \prettyref{fig:criticality_learning}(a)-(b)).
This hyperparameter manifold separates an ordered and a chaotic phase
for which the network signal for different inputs either strongly
correlates or decorrelates. Therefore, this is referred to as the
critical point. Close to criticality the signal can propagate to large
depths and thus network trainability in deep feed-forward neural networks
is improved. This is known as edge-of-chaos initialization \cite{Poole16_3360,Schoenholz17_iclr}
and closely related to the idea of dynamical isometry \cite{Saxe14_iclr,Pennington17_04735,Burkholz19_neurips}.

Due to the gradient response function appearing in the self-consistency
equation \prettyref{eq:CFL_Full_Tilde_Kernel_Propagation} for feature
learning corrections, these fluctuation corrections also propagate
furthest close to criticality (see \prettyref{fig:criticality_learning}(d)).
However, the error signal $\tilde{C}^{(L)}$ \eqref{eq:C_tilde_final}
itself depends non-linearly on both weight variance $g_{l}$ and bias
variance $g_{b}$, becoming largest for small weight variance $g_{l}$
(see \prettyref{fig:criticality_learning}(d)). To get a qualitative
idea, we study the interplay of these two effects at the NNGP, which
corresponds to the initial iteration step of the full self-consistent
solution. The total gradient signal consisting of the error signal
$\tilde{C}^{(L)}$ and the gradient response $\chi^{(l),\leftarrow}$
depends on the network depth (see \prettyref{fig:criticality_learning}(c)).
The product of the two terms in \eqref{eq:backprop_C_tilde_total}
leads to a peak of $\tC^{(1)}$ in the first layer at a weight variance
$g_{l}\simeq g_{f}$ , which is well below the critical value $g_{\text{crit}}$
(see \prettyref{fig:criticality_learning}(d)). Numerical evidence
in \prettyref{fig:criticality_learning}(e) confirms that indeed kernel
adaptation in fully trained networks tends to increase up to around
$g\simeq g_{f}$ when approached from above. Below $g_{f}$ adaptation
suddenly drops, as the network enters the regime of vanishing gradients.
While criticality is known to be not the only relevant criteria for
network training \cite{Bukva23_arxiv}, we are able to explicitly
point out the interaction of criticality with other factors like output
scale.

\subsection{Downscaling of network output enhances feature learning}

Now that we have seen how feature learning corrections result from
the interplay between response function and error signal of the output
kernel, we can ask how we can promote feature learning in deep neural
networks. While the response function depends on the behavior across
layers, the error signal depends solely on the output layer. Reducing
only the weight variance of the output layer $g_{L}$ shrinks the
scale of the output kernel $C^{(L)}$ relative to the target kernel
$YY^{\T}$, thereby directly increasing feature learning corrections
in \prettyref{eq:CFL_Full_Tilde_Kernel_Propagation}.

\begin{figure}[t]
\vspace{0.2in}

\includegraphics{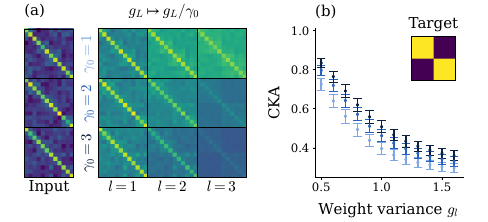}
\centering{}\caption{Increase of feature scale leads to stronger kernel adaption. (a) Network
kernel $C^{(l)}$ across layers $l=1,\,2,\,3$ for different values
of feature scale $\gamma_{0}$. Despite only rescaling the output
layer, feature learning is enhanced in all network layers. Other parameters:
XOR task with $\sigma^{2}=0.4$, $P=12,\,N=200,\,L=3,\,g_{l}=g=0.5$.
(b) CKA between output kernel $C^{(L)}$ and target kernel $YY^{\mathsf{T}}$
for different levels of feature scaling ($\gamma_{0}=1,\,2,\,3$ from
dark to light blue). Larger feature scale consistently leads to stronger
kernel adaptation. Results are averaged over $10$ training data sets
and error bars indicate standard deviation. Other parameters: XOR
task with $\sigma^{2}=0.4$, $P=12,\,L=3$.\label{fig:feature_scaling}}
\vspace{0.2in}
\end{figure}

Previous works \cite{Geiger20_113301,Geiger21_1,Yang20_14522,Yang21_iclr,Bordelon23_114009}
studied how the scaling of the output layer affects the transition
between lazy and feature learning. We here consider the case where
the output weight variance is reduced by a factor $\gamma_{0}$ that
is not extensive in the number of hidden units so that $g_{L}\mapsto g_{L}/\gamma_{0}$.
To understand the effect of such a feature scale $\gamma_{0}$ on
the self-consistency equations \prettyref{eq:CFL_Full_Tilde_Kernel_Propagation},
we derive the dependence of feature learning corrections on the output
kernel $C_{\alpha\beta}^{(L)}\propto g_{L}/\gamma_{0}$ as $\tilde{C}_{\alpha\beta}^{(L)}\stackrel{(\ref{eq:C_tilde_final})\text{ for }\kappa=0}{\propto}\gamma_{0}^{2}+\mathcal{O}(\gamma_{0}).$
From \prettyref{eq:backward_response} follows $\chi^{(L),\leftarrow}\propto g_{L}/\gamma_{0}$,
so that the fluctuation corrections resulting from the conjugate kernel
in the input layer $\tilde{C}_{\alpha\beta}^{(0)}$ increase linearly
with the feature scale
\begin{align}
\tilde{C}_{\alpha\beta}^{(0)} & =\chi^{(L),\leftarrow}\tilde{C}_{\alpha\beta}^{(L)}\propto\gamma_{0}+\mathcal{O}_{\gamma_{0}}(1),\label{eq:feature_scaling}
\end{align}
leading to a stronger adaptation of the network to given training
data. From \prettyref{eq:feature_scaling} follows that gradually
increasing the feature scale $\gamma_{0}$ consistently increases
feature learning in all network layers \prettyref{fig:feature_scaling}(a).
The intuition for this effect is that the reduced scale of the output
kernel $C^{(L)}$ causes the network kernels $C^{(l)}$ to expand
into the direction of the target kernel $YY^{\T}.$ While the interplay
between criticality and weight variance $g_{l}$ for $l<L$ from the
previous subsection stays the same, increasing the feature scale overall
increases feature learning for any weight variance $g_{l}$ \prettyref{fig:feature_scaling}(b).

\section{Discussion}

We here present a new theoretical framework that describes how network
kernels adapt non-linearly to training data, thereby learning features
of a given task. We present two complimentary approaches: For the
proportional limit $P=\alpha N\to\infty$, we employ a large deviation
principle which yields a pair of forward-backward propagation equations
for the maximum a posteriori kernel and its conjugate kernel that
need to be solved self-consistently. We here explore a perturbative
solution in $N^{-1}$, which can be regarded as the limit where $P/N=\alpha\ll1$,
which yet shows reasonable agreement with fully trained networks.
In particular, it correctly predicts the adaptation of the kernel
to the target in deep non-linear networks, in contrast to the kernel
scaling theory for deep linear networks \cite{Li21_031059} or shallow
non-linear networks in a Gaussian equivalence setting \cite{Baglioni24_arxiv}.
In the limit $P=\const$, $N\to\infty$, the solutions of our theory
converge to the NNGP as they should. A complimentary view obtains
qualitatively identical equations for finite-width networks by deriving
kernels from fluctuating auxiliary variables. This view shows that
the network prior comprises a plethora of kernels in the form of a
superposition of Gaussians, a result that holds exactly. When conditioning
on the training data in a Bayesian manner, the Gaussian components
and their associated kernels get reweighed in the network posterior,
yielding a data-dependent maximum a posteriori kernel. The kernel
fluctuations allow the network to sample from many different kernels,
making it more adaptive to data.

In addition to obtaining data-dependent posterior network kernels,
the presented theory allows us to understand driving forces behind
feature learning: We observe an interplay between the response function
of the network and the error signal that is being propagated backwards
through the network by the response function. While being close to
criticality allows the signal to propagate to deep layers \cite{Schoenholz17_iclr},
the error signal itself depends differently on network hyperparameters.
In consequence, kernel adaptation is strongest slightly away from
criticality. Finally, we see how downscaling the network output by
feature scaling increases the error signal, thereby promoting feature
learning in the network.

\paragraph*{Limitations}

For the self-consistency equations to be tractable for non-linear
networks, we approximate them to linear order in the conjugate kernels.
This assumes small corrections relative to the NNGP limit, more specifically
$\alpha=P/N\ll1$. For linear networks, this additional approximation
is not required. By iterating from wider networks to more narrow
networks, we are able to determine kernel corrections for different
network widths. Nevertheless, the here presented approach is strictly
valid only for large $N$, since we use a large deviation principle,
and for non-linear networks is limited to small amounts of training
data relative to the network size $\alpha=P/N\ll1$.

\paragraph*{Outlook}

While the here presented results focus on kernels, we aim to extend
the theoretical framework to study the predictor statistics in the
future. This requires computing non-Gaussian corrections from the
posterior of kernels and determining the interaction between test
samples with training samples. The theoretical framework can be straightforwardly
extended to other network architectures such as RNNs, CNNs, and ResNets,
using the respective network priors \cite{Segadlo22_103401,Garriga18,Fischer23_arxiv}.
Investigating the differences in kernel adaptation for these network
architectures is an interesting question for future work. To study
the effect of noise in input data on feature learning \cite{Lindner23_arxiv},
we plan to include fluctuations of the input kernel in the theoretical
framework. Furthermore, the theoretical framework can be extended
to study feature learning in other network architectures such as transformers,
for which the NNGP is already known \cite{Hron20_4376}. We believe
that the presented theoretical framework constitutes a versatile tool
for studying different aspects of data-dependent kernels and feature
learning.

\section*{Impact Statement}

Engineering of novel technologies in AI continues to supersede our
theoretical understanding of it. Understanding what mechanisms drive
kernel adaptation in neural networks is highly relevant for task-sensitive
hyperparameter optimization (HPO) as it enables informed decisions
about network width, network depth, network initialization etc. In
this work, we draw a novel link between kernel adaptation and criticality,
showing that maximal kernel adaptation happens at prior weight variances
that are significantly different from those predicted by criticality
only.

\section*{Acknowledgements}

We thank Claudia Merger, Itay Lavie, Inbar Seroussi, and Noa Rubin
for helpful discussions. This work was partly supported by the German
Federal Ministry for Education and Research (BMBF Grant 01IS19077A
to Jülich and BMBF Grant 01IS19077B to Aachen) and funded by the Deutsche
Forschungsgemeinschaft (DFG, German Research Foundation) - 368482240/GRK2416
, and the Helmholtz Association Initiative and Networking Fund under
project number SO-092 (Advanced Computing Architectures, ACA). This
work was supported by a fellowship of the German Academic Exchange
Service (DAAD). Open access publication funded by the Deutsche Forschungsgemeinschaft
(DFG, German Research Foundation) -- 491111487. MK would like to
thank the Institute for Advanced Simulation (IAS-6) at Forschungszentrum
Jülich and its directors Markus Diesmann and Sonja Grün for their
hospitality during regular visits.\bibliographystyle{./icml2024}
\bibliography{brain,add_to_brain}
\newpage{}

\appendix
\onecolumn

\part*{Appendix}

\section{Detailed derivation of feature learning theory\label{app:theory}}

In this appendix we present all details of the calculations leading
to the self-consistency equations in \prettyref{sec:theory}. We start
from the network architecture
\begin{align}
h_{\alpha}^{(0)} & =W^{(0)}x_{\alpha}+b^{(0)},\\
h_{\alpha}^{(l)} & =W^{(l)}\phi\left(h_{\alpha}^{(l-1)}\right)+b^{(l)}\quad l=1,\ldots,L,\\
f_{\alpha} & =h_{\alpha}^{(L)},\label{eq:Appendix_NetworkArchitecture}
\end{align}
with Gaussian i.i.d. priors $W_{ij}^{(0)}\stackrel{\text{i.i.d.}}{\sim}\mathcal{N}\left(0,g_{0}/D\right),\,W_{ij}^{(l)}\stackrel{\text{i.i.d.}}{\sim}\mathcal{N}\left(0,g_{l}/N\right),\,W_{i}^{(L)}\stackrel{\text{i.i.d.}}{\sim}\mathcal{N}\left(0,g_{L}/N\right),\,b_{i}^{(l)}\stackrel{\text{i.i.d.}}{\sim}\mathcal{N}(0,g_{b})$
on the network weights and biases. We assume that the network width
$N$ is the same across all layers $l=0,\dots,L$. We condition on
the set of training data consisting of inputs $X=(x_{\alpha})_{\alpha=1,\ldots,P}$
and corresponding labels $Y=(y_{\alpha})_{\alpha=1,\ldots,P}$.

\subsection{Network prior\label{sec:Network-prior}}

With the assumption of sample-wise Gaussian regularization noise $\kappa$,
the prior reads
\begin{equation}
p(Y\vert X)=\int\,\prod_{\alpha=1}^{P}\N(y_{\alpha}|f_{\alpha},\kappa)\,p(f|X)\,df.
\end{equation}
We obtain $p(f\vert X)$ by enforcing the network architecture \prettyref{eq:Appendix_NetworkArchitecture}
using Dirac Delta distributions
\begin{align}
p(f|X) & =\int\mathcal{D}\{W^{(l)},b^{(l)}\}\prod_{\alpha=1}^{P}\delta\left(-f_{\alpha}+W^{(L)}\phi\left(h_{\alpha}^{(L-1)}\right)+b^{(L)}\right)\label{eq:Appendix_Prior_DeltaDistributions}\\
 & \qquad\times\prod_{l=1}^{L-1}\delta\Big(-h_{\alpha}^{(l)}+W^{(l)}\,\phi\left(h_{\alpha}^{(l-1)}\right)+b^{(l)}\Big)\\
 & \qquad\times\delta\left(-h_{\alpha}^{(0)}+W^{(0)}x_{\alpha}+b^{(0)}\right),
\end{align}
where we use the shorthand $\mathcal{D}\{W^{(l)},b^{(l)}\}$ to indicate
the Gaussian measures given by the prior distributions on the weights
and biases. In order to perform the averages, we express the Dirac
Delta distributions using their Fourier transform $\delta(x)=1/(2\pi i)\int_{-i\infty}^{i\infty}\exp(x\tilde{x})\,d\tilde{x}$,
yielding
\begin{align}
\text{\ensuremath{\delta\left(-h_{\alpha k}^{(l)}+\sum_{j=1}^{N}W_{kj}^{(l)}\phi\left(h_{\alpha j}^{(l-1)}\right)+b_{k}^{(l)}\right)}} & =\int_{-i\infty}^{i\infty}\frac{d\tilde{h}_{\alpha k}^{(l)}}{2\pi i}\,\exp\left(-h_{\alpha k}^{(l)}\tilde{h}_{\alpha k}^{(l)}+\tilde{h}_{\alpha k}^{(l)}\sum_{j=1}^{N}W_{kj}^{(l)}\phi\left(h_{\alpha j}^{(l-1)}\right)+\tilde{h}_{\alpha k}^{(l)}b_{k}^{(l)}\right),\label{eq:Appendix_DeltaConstraint_FinalLayer}\\
\text{\ensuremath{\delta\left(-f_{\alpha}+\sum_{j=1}^{N}W_{j}^{(L)}\phi\left(h_{\alpha j}^{(L-1)}\right)+b_{k}^{(L)}\right)}} & =\int_{-i\infty}^{i\infty}\frac{d\tilde{f}_{\alpha}}{2\pi i}\,\exp\left(-f_{\alpha}\tilde{f}_{\alpha}+\tilde{f}_{\alpha}\sum_{j=1}^{N}W_{j}^{(L)}\phi\left(h_{\alpha j}^{(L-1)}\right)+\tilde{f}_{\alpha}b^{(L)}\right).
\end{align}
By doing the Fourier transform, we introduce conjugate variables $\tilde{f}$
for the network output $f$ and $\tilde{h}^{(l)}$ for the layer activations
$h^{(l)}$. Next we perform the averages over network parameters $\Theta=\lbrace W^{(l)},b^{(l)}\rbrace_{l}$
which are i.i.d. Gaussian random variables. In doing so, we identify
the moment generating function (MGF) of these variables; for a Gaussian
it computes to $\langle\exp(j\,x)\rangle_{x\sim\mathcal{N}(0,C)}=\exp\left(C/2\,j^{2}\right)$.
Thus we get for the final layer of the network
\begin{align}
\prod_{\alpha}\left\langle \exp\left(\tilde{f}_{\alpha}\sum_{j=1}^{N}\,W_{j}^{(L)}\phi_{\alpha j}^{(L-1)}+\tilde{f}_{\alpha}b^{(L)}\right)\right\rangle _{W^{(L)},b^{(L)}} & =\left\langle \exp\left(\sum_{\alpha,j}\,\tilde{f}_{\alpha}W_{j}^{(L)}\phi_{\alpha j}^{(L-1)}\right)\right\rangle _{W^{(L)}}\left\langle \exp\left(\sum_{\alpha}\tilde{f}_{\alpha}b^{(L)}\right)\right\rangle _{b^{(L)}},\nonumber \\
 & \overset{\text{MGF}}{=}\exp\left(\frac{g_{L}}{2N}\,\sum_{\alpha\beta}\tilde{f}_{\alpha}\left[\sum_{j=1}^{N}\phi_{\alpha j}^{(L-1)}\phi_{\beta j}^{(L-1)}\right]\tilde{f}_{\beta}\right)\exp\left(\frac{g_{b}}{2}\sum_{\alpha\beta}\tilde{f}_{\alpha}\tilde{f}_{\beta}\right),
\end{align}
with the shorthand $\phi_{\alpha j}^{(l-1)}=\phi\left(h_{\alpha j}^{(l-1)}\right)$.
All subsequent layers yield a similar structure
\begin{align}
\prod_{\alpha}\left\langle \exp\left(\tilde{h}_{\alpha}^{(l)}\sum_{j=1}^{N}\,W_{j}^{(l)}\phi_{\alpha j}^{(l-1)}+\tilde{h}_{\alpha}^{(l)}b^{(l)}\right)\right\rangle _{W^{(l)},b^{(l)}} & =\left\langle \exp\left(\sum_{\alpha,j}\,\tilde{h}_{\alpha}^{(l)}W_{j}^{(l)}\phi_{\alpha j}^{(l-1)}\right)\right\rangle _{W^{(l)}}\left\langle \exp\left(\sum_{\alpha}\tilde{h}_{\alpha}^{(l)}b^{(l)}\right)\right\rangle _{b^{(l)}},\nonumber \\
 & \overset{\text{MGF}}{=}\exp\left(\frac{g_{l}}{2N}\,\sum_{\alpha\beta}\tilde{h}_{\alpha}^{(l)}\left[\phi^{(l-1)}\phi^{(l-1)\T}\right]_{\alpha\beta}\tilde{h}_{\beta}^{(l)}\right)\exp\left(\frac{g_{b}}{2}\sum_{\alpha\beta}\tilde{h}_{\alpha}^{(l)}\tilde{h}_{\beta}^{(l)}\right),
\end{align}
where we introduced $\left[\phi^{(l-1)}\phi^{(l-1)\T}\right]_{\alpha\beta}:=\sum_{j}\phi_{\alpha j}^{(l-1)}\phi_{\beta j}^{(l-1)}.$
The exception is the input layer, which contains the input overlap
matrix $XX^{\mathsf{T}}=\sum_{j=1}^{D}x_{\alpha j}x_{\beta j}$
\begin{align}
\prod_{\alpha}\left\langle \exp\left(\sum_{i,j}\,\tilde{h}_{\alpha i}^{(0)}W_{ij}^{(0)}x_{\alpha j}+\tilde{h}_{\alpha i}^{(0)}b_{i}^{(0)}\right)\right\rangle _{W^{(0)},b^{(0)}} & =\exp\left(\frac{g_{0}}{2D}\,\sum_{\alpha\beta i}\,\tilde{h}_{\alpha i}^{(0)}\big[XX^{\mathsf{T}}\big]{}_{\alpha\beta}\tilde{h}_{\beta i}^{(0)}\right)\exp\left(\frac{g_{b}}{2}\sum_{\alpha\beta i}\,\tilde{h}_{\alpha i}^{(0)}\tilde{h}_{\beta i}^{(0)}\right).
\end{align}
From this we can see that introducing the auxiliary variables
\begin{align}
C_{\alpha\beta}^{(0)} & =\frac{g_{0}}{D}\left(XX^{\mathsf{T}}\right)_{\alpha\beta}+g_{b},\label{eq:Appendix_Auxiliary_Input_Kernel}\\
C_{\alpha\beta}^{(l)} & =\frac{g_{l}}{N}\left[\phi^{(l-1)}\phi^{(l-1)\T}\right]_{\alpha\beta}+g_{b}\quad l=1,\ldots,L,\label{eq:Appendix_Auxiliary_Layer_Kernels}
\end{align}
is beneficial as we can show that
\begin{align*}
\int\mathcal{D}\tilde{h}_{i}^{(l)}\exp\left(-\tilde{h}_{\alpha i}^{(l)}h_{\alpha i}^{(l)}+\frac{1}{2}\,\sum_{\alpha\beta}\tilde{h}_{\alpha i}^{(l)}C_{\alpha\beta}^{(l)}\tilde{h}_{\beta i}^{(l)}\right) & =\mathcal{N}\left(h_{i}^{(l)}\vert0,C_{\alpha\beta}^{(l)}\right)\quad0\le l<L,
\end{align*}
by the Fourier representation of the Gaussian. Likewise one obtains
$\N(f|0,C_{\alpha\beta}^{(L)})$. The form above shows that the $h_{\alpha i}^{(l)}$
are independent across neuron index $i$. Note that the input kernel
$C^{(0)}$ is static for fixed input data sets $X$, whereas all subsequent
auxiliary variables $C^{(l)}$ include network activations $\phi_{\alpha j}^{(L-1)}$
and are hence fluctuating. We now enforce the structure of the fluctuating
auxiliary variables using Dirac Delta distributions
\begin{align}
\delta\left(-C_{\alpha\beta}^{(l)}+\frac{g_{l}}{N}\big[\phi^{(l-1)}\phi^{(l-1)\T}\big]_{\alpha\beta}+g_{b}\right) & =\int_{-i\infty}^{i\infty}\frac{d\tilde{C}_{\alpha\beta}^{(l)}}{2\pi i}\,\exp\left(-\tilde{C}_{\alpha\beta}^{(l)}C_{\alpha\beta}^{(l)}+\tilde{C}_{\alpha\beta}^{(l)}\frac{g_{l}}{N}\big[\phi^{(l-1)}\phi^{(l-1)\T}\big]_{\alpha\beta}+\tilde{C}_{\alpha\beta}^{(l)}g_{b}\right).
\end{align}
Combining all those expressions in \prettyref{eq:Appendix_Prior_DeltaDistributions}
yields $p(f|X)$, which depends on $X$ only through $C^{(0)}$ given
by \eqref{eq:Appendix_Auxiliary_Input_Kernel}
\begin{align}
p(f|X) & \equiv p(f|C^{(0)})\\
 & =\int\mathcal{D}\{\tilde{C},C\}\mathcal{N}\left(f\vert0,C_{\alpha\beta}^{(L)}\right)\left\langle \exp(\mathcal{S}(C,\tilde{C}))\right\rangle _{h},\\
\mathcal{S}(C,\tilde{C}) & =-\sum_{l=1}^{L}\tilde{C}_{\alpha\beta}^{(l)}C_{\alpha\beta}^{(l)}+\tilde{C}_{\alpha\beta}^{(l)}\,\big(\frac{g_{l}}{N}\big[\phi^{(l-1)}\phi^{(l-1)\T}\big]_{\alpha\beta}+g_{b}\big),\nonumber 
\end{align}
where the average $\langle\ldots\rangle_{h}$ indicates the averaging
over the Gaussian distributed hidden states $\mathcal{N}\left(h_{i}^{(l)}\vert0,C_{\alpha\beta}^{(l)}\right)$
and repeated indices $\alpha$, $\beta$ are summed over. As these
distributions are independent for each neuron index $j$, averaging
the third line reduces to
\begin{equation}
\left\langle \exp\left(\tilde{C}_{\alpha\beta}^{(l)}\frac{g_{l}}{N}\sum_{j=1}^{N}\phi_{\alpha j}^{(l-1)}\phi_{\beta j}^{(l-1)}\right)\right\rangle _{\{\mathcal{N}\left(h_{j}^{(l-1)}\vert0,C_{\alpha\beta}^{(l-1)}\right)\}_{j}}\overset{h_{j}^{(l-1)}\,\text{i.i.d. in}\,j}{=}\left\langle \exp\left(\frac{g_{l}}{N}\tilde{C}_{\alpha\beta}^{(l)}\phi_{\alpha}^{(l-1)}\phi_{\beta}^{(l-1)}\right)\right\rangle _{\mathcal{N}\left(h^{(l-1)}\vert0,C_{\alpha\beta}^{(l-1)}\right)}^{N}.
\end{equation}
Overall, this yields
\begin{align}
p(f|X) & =\int\mathcal{D}\{\tilde{C},C\}\mathcal{N}\left(f\vert0,C_{\alpha\beta}^{(L)}\right)\exp\big(-\sum_{l=1}^{L}\tilde{C}_{\alpha\beta}^{(l)}C_{\alpha\beta}^{(l)}+\W(\tC|C)\big),\\
\W(\tC|C) & =\sum_{l=1}^{L}\sum_{\alpha\beta}\tilde{C}_{\alpha\beta}^{(l)}g_{b}+N\,\sum_{l=1}^{L}\ln\,\Big\langle\exp\big(\frac{g_{l}}{N}\tilde{C}_{\alpha\beta}^{(l)}\phi_{\alpha}^{(l-1)}\phi_{\text{\ensuremath{\beta}}}^{(l-1)}\big)\Big\rangle_{\N(0,C^{(l-1)})},\label{eq:CGF_appendix}\\
C_{\alpha\beta}^{(0)} & =\frac{g_{0}}{D}\left(XX^{\T}\right)_{\alpha\beta}+g_{b}.\label{eq:C_0_initial}
\end{align}
As we are interested in the prior $p(Y\vert X)$ and assume Gaussian
i.i.d. noise on the labels, the prior reads
\begin{align*}
p(Y\vert X) & =\int\mathcal{D}\{\tilde{C},C\}\prod_{\alpha}df_{\alpha}\,\N(y_{\alpha}|f_{\alpha},\kappa)\mathcal{N}\left(f_{\alpha}\vert0,C_{\alpha\beta}^{(L)}\right)\exp(-\tr\,\tC^{\T}C+\W(\tC|C)).
\end{align*}
Here we use the same shorthand as in the main text $\tr\,\tC^{\T}C=\sum_{\alpha\beta l}\tC_{\alpha\beta}^{(l)}C_{\alpha\beta}^{(l)}$.
The integral over $f_{\alpha}$ has the form of a convolution of the
normal distribution $\N(y_{\alpha}|f_{\alpha},\kappa)\propto\exp\big(-(y_{\alpha}-f_{\alpha})^{2}/(2\kappa)\big)$
with $\mathcal{N}\left(f_{\alpha}\vert0,C_{\alpha\beta}^{(L)}\right)$,
which amounts to the summation of two random variables $\eta_{\alpha}\stackrel{\text{i.i.d.}}{\sim}\N(0,\kappa)$
and $f_{\alpha}\sim\mathcal{N}\left(f_{\alpha}\vert0,C_{\alpha\beta}^{(L)}\right)$,
so their variances add up to the variance $C_{\alpha\beta}^{(L)}+\kappa\I$.
One therefore obtains the expression for the network prior in \prettyref{eq:network_prior}
\begin{align}
p(Y\vert X) & =\int\mathcal{D}C\,\mathcal{N}\left(Y\vert0,C^{(L)}+\kappa\mathbb{I}\right)p(C),\label{eq:CGF_appendix_brief}\\
p(C) & \coloneqq\int\mathcal{D}\tC\,\exp\big(-\tr\,\tC^{\T}C+\W(\tC|C)\big).\label{eq:def_p_C_appendix}
\end{align}

\subsection{Large deviation approach to network posterior}

When writing the integrals for $p(C)$ given by \eqref{eq:def_p_C_appendix}
we see that the conditional probabilities $p\left(C^{(l)}\vert C^{(l-1)}\right)$
appear naturally as Fourier integrals over the conjugate variable
$\tilde{C}^{(l)}$
\begin{align}
p\left(C^{(l)}\vert C^{(l-1)}\right) & =\int\mathcal{D}\tilde{C}^{(l)}\exp\left(-\tr\,\tilde{C}^{(l)\T}C^{(l)}+\mathcal{W}\left(\tilde{C}^{(l)}\vert C^{(l-1)}\right)\right),\label{eq:p_C_conditional_as_Fourier}\\
\mathcal{W}\left(\tilde{C}^{(l)}\vert C^{(l-1)}\right) & =\tilde{C}^{(l)}g_{b}+N\ln\,\Big\langle\exp\big(\frac{g_{l}}{N}\phi^{(l-1)\T}\tilde{C}^{(l)}\phi^{(l-1)}\Big\rangle_{h^{(l-1)}\sim\N(0,C^{(l-1)})}\quad1\le l<L,\label{eq:Appendix_def_W}
\end{align}
where $C^{(0)}$ is given by \eqref{eq:C_0_initial}. Due to the layerwise
summations in \prettyref{eq:CGF_appendix}, the full expressions $p(C)$
consists of the product of these conditional probabilities $p(C)=p\left(C^{(L)}\vert C^{(L-1)}\right)\cdots p\left(C^{(1)}\vert C^{(0)}\right)$.
Computing the conditional probabilities $p\left(C^{(l)}\vert C^{(l-1)}\right)$
by the Fourier integral in \prettyref{eq:p_C_conditional_as_Fourier}
is intractable for general non-linearities. However, we can write
$\mathcal{W}\left(\tilde{C}^{(l)}\vert C^{(l-1)}\right)$ in the form
\begin{align}
\mathcal{W}\left(\tilde{C}^{(l)}\vert C^{(l-1)}\right) & =N\,\lambda\left(\frac{\tilde{C}^{(l)}}{N}\vert C^{(l-1)}\right),\\
\lambda\left(K\vert C^{(l-1)}\right) & \coloneqq K\,g_{b}+\ln\,\Big\langle\exp\big(g_{l}\phi^{(l-1)\T}K\phi^{(l-1)}\Big\rangle_{\N(0,C^{(l-1)})}.
\end{align}
We observe that $\W$ has the form of a scaled cumulant-generating
function so that the limit $\lim_{N\to\infty}N^{-1}\W(N\,K|C^{(l-1)})=\lambda(K|C^{(l-1)})$
exists trivially. This allows us to utilize the G\"artner-Ellis theorem
to approximate $\ln p\left(C^{(l)}\vert C^{(l-1)}\right)$ for $N\gg1$
(see, e.g., \cite{Touchette09}, i.p. their Appendix C) by a large-deviation
principle as
\begin{align}
-\ln p\left(C^{(l)}\vert C^{(l-1)}\right) & \simeq\sup_{\tilde{C}^{(l)}}\tr\tilde{C}^{(l)\top}C^{(l)}-\mathcal{W}\left(\tilde{C}^{(l)}\vert C^{(l-1)}\right)\label{eq:Appendix_Gamma_Def}\\
 & \eqqcolon\Gamma\left(C^{(l)}\vert C^{(l-1)}\right),\nonumber 
\end{align}
with the rate function $\Gamma\left(C^{(l)}\vert C^{(l-1)}\right)$.
We can hence approximate the full distribution $p(C)$ by
\begin{align}
\ln p(C) & =\ln\left(p\left(C^{(L)}\vert C^{(L-1)}\right)\cdots p\left(C^{(1)}\vert C^{(0)}\right)\right),\\
 & \simeq-\sum_{l=1}^{L}\Gamma\left(C^{(l)}\vert C^{(l-1)}\right)\eqqcolon-\Gamma(C).\label{eq:Gamma_as_sum}
\end{align}
With the rate function we can express the prior $p(Y\vert X)$ as:
\begin{align}
p(Y\vert X) & \simeq\int\mathcal{D}C\mathcal{N}\left(Y\vert0,C^{(L)}+\kappa\mathbb{I}\right)\exp\big(-\Gamma(C)\big),
\end{align}
From the supremum condition \prettyref{eq:Appendix_Gamma_Def} and
by evaluating the integral
\begin{align}
p(Y\vert X) & \simeq\int\mathcal{D}C\exp\big(\mathcal{S}(C)\big),\label{eq:Appendix_Action_S}\\
\S(C) & :=\ln p(C\vert Y)\stackrel{\text{l.d.p.}}{\simeq}\mathcal{S}_{\mathrm{D}}(C^{(L)})-\Gamma(C),\label{eq:action_C_appendix}\\
\mathcal{S}_{\mathrm{D}}(C^{(L)}) & :=-\frac{1}{2}Y^{\top}(C^{(L)}+\kappa\I)^{-1}Y-\frac{1}{2}\ln\det(C^{(L)}+\kappa\I),\label{eq:SD_app}
\end{align}
in a saddle-point approximation in $C^{(l)}$, we obtain the equations
for the network kernels $C^{(l)}$ and the conjugate kernels $\tilde{C}^{(l)}$.

\subsection{Maximum a posteriori network kernels $C^{(l)}$\label{app:Maximum-network-kernels}}

The definition of $\Gamma(C^{(l)}\vert C^{(l-1)})$ \prettyref{eq:Appendix_Gamma_Def}
enforces the supremum in $\tilde{C}^{(l)}$. Hence we require stationarity
in $\tilde{C}^{(l)}$ 
\[
\frac{\partial}{\partial\tilde{C}^{(l)}}\left[\tr\tilde{C}^{(l)\top}C^{(l)}-\mathcal{W}\left(\tilde{C}^{(l)}\vert C^{(l-1)}\right)\right]\overset{!}{=}0
\]
to obtain the supremum as
\begin{align}
C_{\alpha\beta}^{(l)}-\frac{\partial\mathcal{W}}{\partial\tilde{C}_{\alpha\beta}^{(l)}} & =0\\
\rightarrow C_{\alpha\beta}^{(l)} & =g_{l}\left\langle \phi_{\alpha}^{(l-1)}\phi_{\beta}^{(l-1)}\right\rangle _{\mathcal{P}^{(l-1)}}+g_{b}\label{eq:prop_general_theory_appendix}
\end{align}
where we used the form of $\mathcal{W}$ in \prettyref{eq:Appendix_def_W}
and introduced $\left\langle \ldots\right\rangle _{\mathcal{P}^{(l)}}$
to indicate averages over the measure $\mathcal{P}^{(l)}$, which
is given by
\begin{equation}
\left\langle \ldots\right\rangle _{\mathcal{P}^{(l)}}=\frac{\Big\langle\ldots\exp\left(\frac{g_{l}}{N}\phi^{(l)\T}\tilde{C}^{(l+1)}\phi^{(l)}\right)\Big\rangle_{\N(0,C^{(l)})}}{\Big\langle\exp\left(\frac{g_{l}}{N}\phi^{(l)\T}\tilde{C}^{(l+1)}\phi^{(l)}\right)\Big\rangle_{\N(0,C^{(l)})}}.\label{eq:Appendix_Measure_P}
\end{equation}
 This expression corresponds to \ref{eq:prop_general} in the main
text.

\subsection{Self-consistent conjugate kernels $\tilde{C}^{(l)}$\label{app:Maximum-a-posteriori}}

By performing the saddle-point approximation of \prettyref{eq:Appendix_Action_S}
in $C^{(l)}$, we obtain expressions for $\tilde{C}^{(l)}$. We first
need two fundamental properties that follow from the Legendre transform
in the definition of $\Gamma(C^{(l)}|C^{(l-1)})$: The first is the
equation of state
\begin{align}
\frac{\partial}{\partial C^{(l)}}\Gamma(C^{(l)}|C^{(l-1)}) & =\tC^{(l)},\label{eq:eq_state}
\end{align}
which follows because the supremum condition yields
\begin{align*}
\text{} & \ensuremath{\frac{\partial}{\partial C^{(l)}}\sup_{\tC^{(l)}}\,\tr\tC^{(l)\T}C^{(l)}}-\W(\tC^{(l)}|C^{(l-1)})\\
= & \tC^{(l)}+\tr\frac{\partial\tC^{(l)\T}}{\partial C^{(l)}}\,C^{(l)}-\tr\underbrace{\frac{\partial\W(\tC^{(l)}|C^{(l-1)})^{\T}}{\partial\tC^{(l)}}}_{\equiv C^{(l)\T}}\frac{\partial\tC^{(l)}}{\partial C^{(l)}}
\end{align*}
so that the latter two terms cancel each other.

The second fundamental property of the Legendre transform applies
to the derivative by $C^{(l-1)}$, which here plays the role of a
parameter, for which holds
\begin{align}
\frac{\partial}{\partial C^{(l-1)}}\,\Gamma(C^{(l)}|C^{(l-1)}) & =-\frac{\partial\W(\tC^{(l)}|C^{(l-1)})}{\partial C^{(l-1)}},\label{eq:param_dep_gamma}
\end{align}
again, because of the supremum condition
\begin{align*}
 & \ensuremath{\frac{\partial}{\partial C^{(l-1)}}\sup_{\tC^{(l)}}\,\tr\tC^{(l)\T}C^{(l)}}-\W(\tC^{(l)}|C^{(l-1)})\\
 & =\tr\frac{\partial\tC^{(l)\T}}{\partial C^{(l-1)}}C^{(l)}-\tr\underbrace{\frac{\partial\W^{\T}}{\partial\tC^{(l)}}}_{\equiv C^{(l)}}\,\frac{\partial\tC^{(l)}}{\partial C^{(l-1)}}-\frac{\partial\W}{\partial C^{(l-1)}},
\end{align*}
the first two terms on the right hand side cancel. The stationary
points of $\eqref{eq:action_C_appendix}$ for $1\le l<L$ then follow
with the explicit form of $\Gamma(C)=\sum_{l=1}^{L}\Gamma\left(C^{(l)}\vert C^{(l-1)}\right)$
from \eqref{eq:Gamma_as_sum} as
\begin{align}
0 & \stackrel{!}{=}\frac{\partial\S(C)}{\partial C^{(l)}}=\frac{\partial}{\partial C^{(l)}}\sum_{l=1}^{L}\Gamma\left(C^{(l)}\vert C^{(l-1)}\right)\label{eq:stat_intermediate}\\
 & =\frac{\partial\Gamma(C^{(l)}|C^{(l-1)})}{\partial C^{(l)}}+\frac{\partial\Gamma(C^{(l+1)}|C^{(l)})}{\partial C^{(l)}}\nonumber \\
 & \stackrel{(\ref{eq:eq_state}),(\ref{eq:param_dep_gamma})}{=}\tC^{(l)}-\frac{\partial\W(\tC^{(l+1)}|C^{(l)})}{\partial C^{(l)}}.\nonumber 
\end{align}
Using the definition of $\mathcal{W}$, we compute
\begin{align}
\frac{\partial}{\partial C_{\alpha\beta}^{(l)}}\,\W(\tC^{(l+1)}|C^{(l)}) & =N\,\frac{\frac{\partial}{\partial C_{\alpha\beta}^{(l)}}\Big\langle\exp\big(\frac{g_{l+1}}{N}\phi^{(l)\T}\tC^{(l+1)}\phi^{(l)}\big)\Big\rangle_{\N(0,C^{(l)})}}{\Big\langle\exp\big(\frac{g_{l+1}}{N}\phi^{(l)\T}\tC^{(l+1)}\phi^{(l)}\big)\Big\rangle_{\N(0,C^{(l)})}}
\end{align}
by Price's theorem (see \prettyref{app:Generalization-of-Price}):
the derivative in the numerator is
\begin{align}
\frac{\partial}{\partial C_{\alpha\beta}^{(l)}}\Big\langle\exp\left(\frac{g_{l+1}}{N}\phi^{(l)\T}\tC^{(l+1)}\phi^{(l)}\right)\Big\rangle_{h^{(l)}\sim\N(0,C^{(l)})} & \overset{(\ref{eq:price_general-2})}{=}\frac{1}{2}\,\Big\langle\frac{\partial}{\partial h_{\alpha}^{(l)}\partial h_{\beta}^{(l)}}\,\exp\left(\frac{g_{l+1}}{N}\sum_{\gamma\delta}\phi_{\gamma}^{(l)}\tC_{\gamma\delta}^{(l+1)}\phi_{\delta}^{(l)}\right)\Big\rangle_{h^{(l)}\sim\N(0,C^{(l)})},\label{eq:dW_by_dC}
\end{align}
which yields with \eqref{eq:stat_intermediate} for $\tilde{C}^{(l)}$
with $1\le l<L$
\begin{align}
\tilde{C}_{\alpha\beta}^{(l)}=\frac{\partial\W(\tC^{(l+1)}|C^{(l)})}{\partial C_{\alpha\beta}^{(l)}} & =g_{l+1}\,\left[\left\langle \left(\phi_{\alpha}^{(l)}\right)^{\prime}\left(\phi_{\beta}^{(l)}\right)^{\prime}\right\rangle _{\mathcal{P}^{(l)}}\,\tC_{\alpha\beta}^{(l+1)}+\delta_{\alpha\beta}\sum_{\gamma}\left\langle \phi_{\gamma}^{(l)}\left(\phi_{\alpha}^{(l)}\right)^{\prime\prime}\right\rangle _{\mathcal{P}^{(l)}}\tilde{C}_{\gamma\alpha}^{(l+1)}\right]\nonumber \\
 & \quad+2\frac{g_{l+1}^{2}}{N}\,\sum_{\gamma,\delta}\tC_{\alpha\gamma}^{(l+1)}\,\tC_{\beta\delta}^{(l+1)}\,\left\langle \left(\phi_{\alpha}^{(l)}\right)^{\prime}\phi_{\gamma}^{(l)}\,\left(\phi_{\beta}^{(l)}\right)^{\prime}\phi_{\delta}^{(l)}\right\rangle _{\mathcal{P}^{(l)}}\,,\label{eq:dW_dC}
\end{align}
with measure $\mathcal{P}^{(l)}$ as defined in \prettyref{eq:Appendix_Measure_P}.
In the main text in \eqref{eq:CFL_Full_Tilde_Kernel_Propagation},
we only keep terms $\propto\tC$ on the right hand sides of \eqref{eq:dW_dC}.
In the case $l=L$, we additionally need to consider the data-term
in \prettyref{eq:Appendix_Action_S}, yielding
\begin{align}
0 & \stackrel{!}{=}\frac{\partial\S(C)}{\partial C_{\alpha\beta}^{(L)}}=\frac{\partial}{\partial C_{\alpha\beta}^{(L)}}\big(\mathcal{S}_{\mathrm{D}}(C^{(L)})-\Gamma(C^{(L)}|C^{(L-1)})\big)\\
 & \stackrel{(\ref{eq:eq_state})}{=}\frac{1}{2}\left((C^{(L)}+\kappa\I)^{-1}YY^{\T}(C^{(L)}+\kappa\I)^{-1}-(C^{(L)}+\kappa\I)^{-1}\right)-\tC^{(L)},
\end{align}
which is the result \eqref{eq:C_tilde_final} in the main text. To
compute $\partial\mathcal{S}_{\mathrm{D}}(C^{(L)})/\partial C^{(L)}$
with \eqref{eq:SD_app} we used the matrix derivatives
\begin{align}
\frac{\partial\ln\det(C)}{\partial C_{\alpha\beta}} & =\big[C\big]_{\alpha\beta}^{-1},\\
\frac{\partial\big[C\big]_{\gamma\delta}^{-1}}{\partial C_{\alpha\beta}} & =-\big[C\big]_{\gamma\alpha}^{-1}\big[C\big]_{\beta\delta}^{-1},
\end{align}
yielding
\begin{equation}
\tC^{(L)}=\frac{1}{2}(C^{(L)}+\kappa\I)^{-1}(YY^{\T})(C^{(L)}+\kappa\I)^{-1}-\frac{1}{2}(C^{(L)}+\kappa\I)^{-1}.\label{eq:C_tilde_final_layer}
\end{equation}

\subsection{Relation between conjugate kernel and discrepancy\label{app:conj_kernel_discrepancy}}

To understand the meaning of the conjugate kernel $\tilde{C}$, we
generalize the regularization $\kappa\I$ with a generic covariance
matrix $K_{\alpha\beta}$ in \eqref{eq:CGF_appendix_brief}
\begin{align}
p(Y|X,K):= & \int\D C\,\int\D f\,\N(Y|f,K)\,\N(f|0,C^{(L)})\,p(C),\label{eq:p_z_K}
\end{align}
which shows that, given $C^{(L)}$, the statistics of $Y$ is a convolution
of two centered Gaussian distributions with covariances $C^{(L)}$
and $K$, respectively. In large deviation theory, this yields the
action
\begin{align}
S(C|K) & =-\frac{1}{2}y_{\alpha}\big[C^{(L)}+K\big]_{\alpha\beta}^{-1}y_{\beta}-\frac{1}{2}\ln\det(C+K)-\Gamma(C).\label{eq:action_K}
\end{align}
Writing \eqref{eq:p_z_K} explicitly
\begin{align*}
p(Y|X,K)=\frac{1}{(2\pi)^{\frac{M}{2}}\,(\det K)^{\frac{1}{2}}}\,\int\D C\,\int\D f\,\exp\big(-\frac{1}{2}(y_{\alpha}-f_{\alpha})\,\big[K^{-1}\big]_{\alpha\beta}\,(y_{\beta}-f_{\beta})\big)\,\N(f|0,C^{(L)})\,p(C),
\end{align*}
we may use $K^{-1}$ as a bi-linear source term so that we obtain
within the MAP approximation for the kernels $C$, the second moment
of the discrepancies as
\begin{align}
-\frac{1}{2}\,\langle(y_{\alpha}-f_{\alpha})\,(y_{\beta}-f_{\beta})\rangle & =\frac{\partial}{\partial[K^{-1}]_{\alpha\beta}}\,\Big(\ln p(Y|X,K)-\frac{1}{2}\det K^{-1}\Big)\Big|_{K=\kappa\I}\label{eq:C_tilde_discrepancy}\\
 & \stackrel{\text{MAP}}{\simeq}\frac{\partial}{\partial[K^{-1}]_{\alpha\beta}}S(C|K)+\underbrace{\frac{\partial S}{\partial C}}_{=0}\,\frac{\partial C}{\partial(K^{-1})_{\alpha\beta}}-\frac{1}{2}K\,\Big|_{K=\kappa\I}\nonumber \\
 & \stackrel{\eqref{eq:action_K}}{=}\Big[-\frac{1}{2}K\,\big[C^{(L)}+K\big]^{-1}\,YY^{\T}\,\big[C^{(L)}+K\big]^{-1}\,K+\frac{1}{2}\,K\,(C^{(L)}+K)^{-1}\,K\Big]_{\alpha\beta}-\frac{1}{2}K\Big|_{K=\kappa\I}\nonumber \\
 & =\Big[-\frac{1}{2}\kappa^{2}\,\big[C^{(L)}+\kappa\I\big]^{-1}\,(YY^{\T})\,\big[C^{(L)}+\kappa\I\big]^{-1}+\frac{1}{2}\,\kappa^{2}\,(C^{\ast}+\kappa\I)^{-1}-\frac{1}{2}\kappa\I\Big]_{\alpha\beta}\nonumber \\
 & \stackrel{\eqref{eq:C_tilde_final_layer}}{=}-\kappa^{2}\tilde{C}_{\alpha\beta}^{\ast}-\frac{1}{2}\kappa\delta_{\alpha\beta},\nonumber 
\end{align}
where we used that $\partial[K^{-1}]_{\gamma\delta}/\partial K_{\alpha\beta}=-K_{\gamma\alpha}^{-1}\,K_{\beta\delta}^{-1}$
and by symmetry $\partial K_{\gamma\delta}/\partial[K^{-1}]_{\alpha\beta}=-K_{\gamma\alpha}\,K_{\beta\delta}$.
This may also be rewritten as
\begin{align}
\langle\Delta_{\alpha}\,\Delta_{\beta}\rangle & =2\kappa^{2}\tilde{C}_{\alpha\beta}^{(L)}+\kappa\delta_{\alpha\beta},\label{eq:exp_discrepancy_square_tilde_C}\\
\Delta_{\alpha} & =y_{\alpha}-f_{\alpha},\nonumber 
\end{align}
so that the expected training error is
\begin{align}
\langle\cL\rangle:= & \frac{1}{2}\tr\,\langle\Delta\,\Delta\rangle\label{eq:avg_training_loss}\\
= & \kappa^{2}\,\tr\tilde{C}^{(L)}+\frac{1}{2}\kappa P.\nonumber 
\end{align}
Expressions \eqref{eq:exp_discrepancy_square_tilde_C} and \eqref{eq:avg_training_loss}
show that the conjugate kernel $\tC$ is, apart from the diagonal
term, given by the expected squared discrepancies $\Delta$ between
target and network output.

\section{Relation between feature learning and fluctuation corrections\label{app:AppendixFluctuations}}

We here show that the shift of the saddle point of $C^{(l)}$ by conditioning
on the training data can be regarded as accounting for fluctuation
corrections for the auxiliary variables $C$ and $\tilde{C}$ around
the reference point, which is the NNGP. As opposed to the rigorous
approach of the main text that is based on a large deviation principle,
we here obtain the result from a perspective of field theory. To this
end, we rewrite the network prior \eqref{eq:network_prior} as an
integral over the pair of fields $(C,\tilde{C}$) as in \eqref{eq:CGF_appendix_brief}
\begin{align}
p(Y|X) & =\int DC\int D\tC\,\N(Y|0,C^{(L)}+\kappa\I)\,\exp\left(S(C,\tC)\right),\label{eq: prior_with_C_aux}\\
\mathcal{S}(C,\tilde{C}) & :=-\tr\,\tC^{\T}C+\mathcal{W}(\tilde{C}|C),\nonumber 
\end{align}
with cumulant-generating function $\W$ given by \eqref{eq:cum_W}.
Adding the normal distribution in \eqref{eq: prior_with_C_aux} as
$\exp(\mathcal{S}_{\mathrm{D}})$ one has a joint measure for the
pair of variables $(C,\tC)$ which is, up to normalization, given
by
\begin{align*}
(C,\tC) & \sim\exp\big(\S(C,\tC)+\mathcal{S}_{\mathrm{D}}(C^{(L)}|Y)\big)\\
\mathcal{S}_{\mathrm{D}}(C^{(L)}|Y) & :=\ln\N(Y|0,C^{(L)}+\kappa\I)\\
 & \equiv-\frac{1}{2}Y^{\T}(C^{(L)}+\kappa\I)^{-1}\,Y-\frac{1}{2}\,\ln\det(C^{(L)}+\kappa\I).
\end{align*}
Now we will expand $S(C,\tC)$ around its saddle point, which, with
regard to $\tC^{(l)}$ for $1\le l\le L$ yields
\begin{align}
0\stackrel{!}{=}\frac{\partial\S}{\partial\tC_{\alpha\beta}^{(l)}} & =-C_{\alpha\beta}^{(l)}+\frac{\partial\W}{\partial\tC_{\alpha\beta}^{(l)}}\label{eq:prop_general_appendix}\\
 & =-C_{\alpha\beta}^{(l)}+g_{l}\,\left\langle \phi_{\alpha}^{(l-1)}\phi_{\beta}^{(l-1)}\right\rangle _{\cP^{(l-1)}}+g_{b},\nonumber 
\end{align}
and the initial value $C^{(0)}$ given by \eqref{eq:C_0} for $l=0$.
This result is of course identical to \eqref{eq:prop_general}, as
it corresponds to the supremum condition in \eqref{eq:rate_function-1}.
The second saddle point equation for $C^{(L)}$ yields
\begin{align}
0\stackrel{!}{=}\frac{\partial\S}{\partial C_{\alpha\beta}^{(L)}} & =-\tC_{\alpha\beta}^{(L)},
\end{align}
and for $1\le l<L$ the equation is given by \eqref{eq:CFL_Full_Tilde_Kernel_Propagation}
of the main text. Together this shows by induction that the stationary
point is $\tC_{\ast}^{(1\le l\le L)}\equiv0$; so the measure $\cP^{(l-1)}$
appearing in \eqref{eq:prop_general_appendix} reduces to a Gaussian
$\N(0,C^{(l-1)})$ and the propagation of $C^{(l)}$ over layers becomes
the NNGP, whose solution we will denote as $C_{\ast}^{(l)}$ and $\tC_{\ast}^{(l)}\equiv0$.

Computing the next-to-leading-order in $N^{-1}$, we need the Hessian
of the action $\S$, evaluated at the NNGP saddle point, which is
\begin{align}
&\mathcal{S}_{(\alpha\beta)(\gamma\delta)}^{(2)(l,m)}\Big|_{C_{\ast},\tC\equiv0} =\left(\begin{array}{cc}
\frac{\partial^{2}\mathcal{S}}{\partial C_{\alpha\beta}^{(l)}\partial C_{\gamma\delta}^{(m)}} & \frac{\partial^{2}\mathcal{S}}{\partial C_{\alpha\beta}^{(l)}\partial\tilde{C}_{\gamma\delta}^{(m)}}\\
\frac{\partial^{2}\mathcal{S}}{\partial\tilde{C}_{\alpha\beta}^{(l)}\partial C_{\gamma\delta}^{(m)}} & \frac{\partial^{2}\mathcal{S}}{\partial\tilde{C}_{\alpha\beta}^{(l)}\partial\tilde{C}_{\gamma\delta}^{(m)}}
\end{array}\right)\label{eq:Hessian}\\
 & =\left(\begin{array}{cc}
0 & -\delta_{l,m}\,\delta_{(\alpha\beta),(\gamma\delta)}+\delta_{m-1,l}\,g_{m}\,\frac{\partial\langle\phi_{\gamma}^{(m-1)}\phi_{\delta}^{(m-1)}\rangle_{\N(0,C_{\ast}^{(l)})}}{\partial C_{\alpha\beta}^{(l)}}\\
-\delta_{l,m}\,\delta_{(\alpha\beta),(\gamma\delta)}+\delta_{l-1,m}\,g_{l}\,\frac{\partial\langle\phi_{\alpha}^{(l-1)}\phi_{\beta}^{(l-1)}\rangle_{\N(0,C_{\ast}^{(l)})}}{\partial C_{\gamma\delta}^{(m)}} & \quad\delta_{l,m}\,g_{l}^{2}\,\langle\phi_{\alpha}^{(l-1)}\phi_{\beta}^{(l-1)},\phi_{\gamma}^{(l-1)}\phi_{\delta}^{(l-1)}\rangle_{c,\,\N(0,C_{\ast}^{(l)})}
\end{array}\right),\nonumber 
\end{align}
where we used that $\W(0|C)\equiv1\quad\forall C$, so that its derivative
in the upper left element vanishes and all expectation values $\langle\ldots\rangle_{\N(0,C_{\ast}^{(l)})}$
are with regard to the Gaussian measure of the NNGP. Since the expectation
values of products of $\phi^{(l)}$ only depend on the value of $C^{(l)}$
in the same layer, the non-trivial terms in the off-diagonal elements
are proportional to Kronecker symbols $\delta_{m-1,l}$ (upper right)
and $\delta_{l-1,m}$ (lower left). The lower right element contains
the connected two-point correlation function $\langle f,g\rangle_{c}:=\langle fg\rangle-\langle f\rangle\langle g\rangle$,
coming from the second derivative by $\tC$; derivatives by $\tC^{(l)}$
and $\tC^{(m)}$ with $l\neq m$ vanish, because the cumulant-generating
function \eqref{eq:cum_W} decomposes into a sum of cumulant-generating
functions across layers, showing their statistical independence, so
that connected correlations across layers vanish.

Within this expansion, the network prior \eqref{eq: prior_with_C_aux}
takes the form
\begin{align}
p(Y|X) & \simeq\int D\delta C\,\int D\delta\tC\,\exp\Big(\frac{1}{2}(\delta C,\delta\tC)^{\T}\mathcal{S}^{(2)}\,(\delta C,\delta\tC)+\mathcal{S}_{\mathrm{D}}(C_{\ast}^{(L)}+\delta C^{(L)}|Y)\Big)
\end{align}
because for $\tC_{\ast}\equiv0$ the zeroth order Taylor term $\S(C_{\ast},0)\equiv0$
and also the linear term vanishes, because $C_{\ast}$ has been chosen
as the stationary point. We may now study the influence of $\mathcal{S}_{\mathrm{D}}$
on the saddle point of $\delta C$ and $\delta\tC$. This term only
affects the saddle point through its affect on $\delta C^{(L)}$,
namely like a source term $\tr J^{\T}\,\delta C^{(L)}$ with
\begin{align}
J_{\alpha\beta} & :=\frac{\partial\mathcal{S}_{\mathrm{D}}}{\partial C_{\alpha\beta}^{(L)}}.
\end{align}
So the saddle point equation for the shift $(\delta C,\delta\tC)$
of the saddle points reads
\begin{align}
\Big[\mathcal{S}^{(2)}\,\left(\begin{array}{c}
\delta C\\
\delta\tC
\end{array}\right)\Big]^{(l)} & +\left(\begin{array}{c}
J\\
0
\end{array}\right)\,\delta_{l\,L}=0,\label{eq:implicit_saddle_correction}
\end{align}
Written explicitly with help of the Hessian \eqref{eq:Hessian}, the
first line of \eqref{eq:implicit_saddle_correction} therefore reads
\begin{align}
-\delta\tC_{\alpha\beta}^{(l)} & +g_{l+1}\,\sum_{\gamma\delta}\,\frac{\partial\langle\phi_{\gamma}^{(l)}\phi_{\delta}^{(l)}\rangle_{\N(0,C^{(l)})}}{\partial C_{\alpha\beta}^{(l)}}\,\delta\tilde{C}_{\gamma\delta}^{(l+1)}+\delta_{l,L}\,J_{\alpha\beta}=0.\label{eq:lin_response_C_tilde}
\end{align}
Employing Price's theorem \eqref{eq:price_general-2} one has
\begin{align*}
 & \sum_{\gamma\delta}\,\frac{\partial\langle\phi_{\gamma}^{(l)}\phi_{\delta}^{(l)}\rangle_{\N(0,C^{(l)})}}{\partial C_{\alpha\beta}^{(l)}}\,\delta\tilde{C}_{\gamma\delta}^{(l+1)}\\
 & =\frac{1}{2}\,\sum_{\gamma\delta}\,\big\langle\frac{\partial}{\partial h_{\alpha}^{(l)}}\frac{\partial}{\partial h_{\beta}^{(l)}}\,\phi_{\gamma}^{(l)}\phi_{\delta}^{(l)}\big\rangle_{\N(0,C^{(l)})}\,\delta\tilde{C}_{\gamma\delta}^{(l+1)}\\
 & =\big\langle\big(\phi_{\alpha}^{(l)}\big)^{\prime}\big(\phi_{\beta}^{(l)}\big)^{\prime}\big\rangle_{\N(0,C^{(l)})}\,\delta\tilde{C}_{\alpha\beta}^{(l+1)}\\
 & +\delta_{\alpha\beta}\,\sum_{\gamma}\,\big\langle\big(\phi_{\alpha}^{(l)}\big)^{\prime\prime}\big(\phi_{\gamma}^{(l)}\big)\big\rangle_{\N(0,C^{(l)})}\,\delta\tilde{C}_{\alpha\gamma}^{(l+1)},
\end{align*}
where we used that $\tC$ and $C$ are both symmetric. Inserted into
\eqref{eq:lin_response_C_tilde}, we obtain, to linear order in $\tC$,
the same propagation equation as stated in \eqref{eq:CFL_Full_Tilde_Kernel_Propagation}.

The second line of \eqref{eq:implicit_saddle_correction} reads explicitly
\begin{align*}
-\delta C_{\alpha\beta}^{(l)}+g_{l}\,\sum_{\gamma\delta}\,\frac{\partial\langle\phi_{\alpha}^{(l-1)}\phi_{\beta}^{(l-1)}\rangle_{\N(0,C^{(l-1)})}}{\partial C_{\gamma\delta}^{(l-1)}}\,\delta C_{\gamma\delta}^{(l-1)}+g_{l}^{2}\,\sum_{\gamma\delta}\,\langle\phi_{\alpha}^{(l-1)}\phi_{\beta}^{(l-1)},\phi_{\gamma}^{(l-1)}\phi_{\delta}^{(l-1)}\rangle_{c,\,\N(0,C^{(l-1)})}\,\delta\tilde{C}_{\gamma\delta}^{(l)} & =0.
\end{align*}
Rewritten, this reads
\begin{align*}
\delta C_{\alpha\beta}^{(l)} & =g_{l}\,\sum_{\gamma\delta}\frac{\partial\langle\phi_{\alpha}^{(l-1)}\phi_{\beta}^{(l-1)}\rangle_{\N(0,C^{(l-1)})}}{\partial C_{\gamma\delta}^{(l-1)}}\,\delta C_{\gamma\delta}^{(l-1)}+g_{l}^{2}\,\sum_{\gamma\delta}V_{\alpha\beta,\gamma\delta}^{(l-1)}\,\delta\tilde{C}_{\gamma\delta}^{(l)},
\end{align*}
where we used the definition $V_{\alpha\beta,\gamma\delta}^{(l-1)}\equiv\langle\phi_{\alpha}^{(l-1)}\phi_{\beta}^{(l-1)},\phi_{\gamma}^{(l-1)}\phi_{\delta}^{(l-1)}\rangle_{c,\,\N(0,C^{(l-1)})}$
from \eqref{eq:perturbative_forward}. The first term on the right
hand side yields the linear correction to $\delta C$ due to the shift
$\delta C$ in \eqref{eq:prop_general_appendix} and the second term
is identical to the correction from $\delta\tC$ in \eqref{eq:perturbative_forward}.
This shows that the self-consistency equations derived in the main
text, up to linear order in $\delta C$, are identical to taking fluctuations
of $C$ up to Gaussian order into account. This allows us to link
points where these fluctuations are large, critical points, to feature
learning.

\section{Deep linear networks\label{app:linear_net}}

In this section we consider the special case of a deep linear network
to make the connection to previous works \cite{Li21_031059,ZavatoneVeth22_064118,Yang23_39380}.
This case allows us to obtain closed-form expressions for both, the
forward iteration equation \eqref{eq:prop_general} and the backward
equation \eqref{eq:CFL_Full_Tilde_Kernel_Propagation}; these expressions
in particular do not require us to apply a perturbative treatment
but only rest on the use of the large deviation principle.

We also show here that in the case of a linear network, our action
\eqref{eq:action_C}, valid for general non-linear networks, reduces
to the one by \cite{Yang23_39380}, their Eq. (1), derived for deep
kernel machines. This new result has three main implications: 
\begin{enumerate}
\item It shows that in the proportional limit $P=\alpha\,N$ considered
here, deep linear networks reduce to deep kernel machines.
\item The here found iterative forward-backward equations may also be used
to determine the MAP estimate for the kernels in deep linear neural
networks and in deep kernel machines.
\item It allows us to provide the alternative view of kernel adaptation
generated by kernel fluctuations, as outlined in \eqref{sec:criticality_learning},
also for deep linear networks and deep kernel machines; this point
is useful, because it shows how the corrections found in the proportional
$P=\alpha\,N$, $N\to\infty$, apply to networks at finite size.
\end{enumerate}
The derivation here will follow along the same steps as in the general
non-linear case in Appendix \prettyref{app:theory}. We will here
only point out the important differences. The setting of a deep linear
neural network here is \eqref{eq:Appendix_NetworkArchitecture},
only replacing the activation function by the identity $\phi=\mathrm{id}$.
This change only affects the mapping for the hidden layers which are
now
\begin{align*}
h_{\alpha}^{(l)} & =W^{(l)}\,h_{\alpha}^{(l-1)}+b^{(l)}\quad l=1,\ldots,L.
\end{align*}
Moreover, we use the same prior distributions for all parameters as
in the general case. Following the same steps as in \prettyref{sec:Network-prior},
the network prior corresponding to \prettyref{eq:CGF_appendix} of
the non-linear network takes the following form for the linear network
\begin{align}
p(f|X) & =\int\mathcal{D}\{\tilde{C},C\}\,\mathcal{N}\left(f\vert0,C_{\alpha\beta}^{(L)}\right)\exp\big(-\sum_{l=1}^{L}\tilde{C}_{\alpha\beta}^{(l)}C_{\alpha\beta}^{(l)}+\W(\tC|C)\big),\\
\W(\tC|C) & =\sum_{l=1}^{L}\sum_{\alpha\beta}\tilde{C}_{\alpha\beta}^{(l)}g_{b}+N\,\sum_{l=1}^{L}\ln\,\Big\langle\exp\big(\frac{g_{l}}{N}\tilde{C}_{\alpha\beta}^{(l)}h_{\alpha}^{(l-1)}h_{\text{\ensuremath{\beta}}}^{(l-1)}\big)\Big\rangle_{\N(0,C^{(l-1)})},\label{eq:CGF_appendix_deep_linear}\\
C_{\alpha\beta}^{(0)} & =\frac{g_{0}}{D}\left(XX^{\T}\right)_{\alpha\beta}+g_{b}.
\end{align}
In contrast to the non-linear case, here the expectation value in
the definition of $\W(\tC|C)$ is a Gaussian integral with the closed-form
solution
\begin{align}
\W(\tC|C) & =\sum_{l=1}^{L}\W(\tC^{(l)}|C^{(l-1)})\label{eq:W_linear}\\
\W(\tC^{(l)}|C^{(l-1)}):= & \sum_{\alpha\beta}\tilde{C}_{\alpha\beta}^{(l)}g_{b}+N\,\ln\,\big\langle\exp\big(\frac{g_{l}}{N}\,\tC_{\alpha\beta}^{(l)}h_{\alpha}^{(l-1)}h_{\beta}^{(l-1)}\big)\big\rangle_{\N(0,C^{(l-1)})}\nonumber \\
= & \sum_{\alpha\beta}\tilde{C}_{\alpha\beta}^{(l)}g_{b}-\frac{N}{2}\,\ln\det\,\big([C^{(l-1)}]^{-1}-2\frac{g_{l}}{N}\,\tC^{(l)}\big)-\frac{N}{2}\,\ln\det(C^{(l-1)}).\nonumber 
\end{align}
Since this cumulant-generating function has the scaling form so that
the limit $\lambda(k):=\lim_{N\to\infty}\W(Nk)/N$ exists trivially,
we may employ the G\"artner-Ellis theorem to approximate the conditional
probabilities $p(C^{(l)}|C^{(l-1)})$ (cf. \prettyref{eq:p_C_conditional_as_Fourier})
by a rate function $\Gamma$
\begin{align}
-\ln p\left(C^{(l)}\vert C^{(l-1)}\right) & :=-\int\mathcal{D}\tilde{C}^{(l)}\exp\left(-\tr\,\tilde{C}^{(l)\T}C^{(l)}+\mathcal{W}\left(\tilde{C}^{(l)}\vert C^{(l-1)}\right)\right)\nonumber \\
 & \stackrel{\text{l.d.p.}}{\simeq}\Gamma(C^{(l)}|C^{(l-1)})\label{eq:gamma_lin_general}\\
 & :=\sup_{\tilde{C}^{(l)}}\tr\tilde{C}^{(l)\top}C^{(l)}-\mathcal{W}\left(\tilde{C}^{(l)}\vert C^{(l-1)}\right)\nonumber \\
 & =\frac{N}{2g_{l}}\,\tr\big([C^{(l-1)}]^{-1}(C^{(l)}-g_{b})\big)-\frac{N}{2}\,\ln\det\,\big(C^{(l)}-g_{b}\big)+\frac{N}{2}\,\ln\det(C^{(l-1)})+\const\,,\nonumber 
\end{align}
where we dropped terms that are constant in the $C$ and used that
the supremum condition in the penultimate line evaluates to
\begin{align}
0 & \stackrel{!}{=}\frac{\partial}{\partial\tC_{\alpha\beta}^{(l)}}\Big(\tr\tilde{C}^{(l)\top}\tilde{C}^{(l)}-\mathcal{W}\left(\tilde{C}^{(l)}\vert C^{(l-1)}\right)\Big)\label{eq:forward_linear}\\
 & =C_{\alpha\beta}^{(l)}-g_{b}-g_{l}\,\big([C^{(l-1)}]^{-1}-2\frac{g_{l}}{N}\,\tC^{(l)}\big)_{\alpha\beta}^{-1}.\nonumber 
\end{align}
This yields the equation to propagate the kernels $C$ forward through
the network (corresponding to \prettyref{eq:prop_general} and \prettyref{eq:prop_general_theory_appendix}
in the non-linear case), which, for $\tC=0$, again reduces to the
NNGP result as expected. Solved for $\tC$ the latter yields
\begin{align*}
\tC^{(l)} & =\frac{N}{2}\,\big([g_{l}C^{(l-1)}]^{-1}-[C^{(l)}-g_{b}]^{-1}\big),
\end{align*}
which, inserted into the penultimate line of \prettyref{eq:gamma_lin_general}
yields the last line there.

At this point we are able to make the connection to deep kernel machines
studied in \cite{Yang23_39380}. The action for the kernels $C$,
corresponding to \prettyref{eq:action_C} in the non-linear case,
in the case of the linear networks with \prettyref{eq:CGF_appendix_deep_linear}
and \prettyref{eq:gamma_lin_general} takes the form
\begin{align}
\S(C) & :=\ln p(C|Y)\stackrel{\text{l.d.p.}}{\simeq}\mathcal{S}_{\mathrm{D}}(C^{(L)})-\Gamma(C)+\circ\,,\label{eq:action_linear}\\
\Gamma(C) & =\sum_{l=1}^{L}\Gamma(C^{(l)}|C^{(l-1)}).\nonumber 
\end{align}
The specific form of the rate function $\Gamma(C^{(l)}|C^{(l-1)})$
in \prettyref{eq:gamma_lin_general} is that of a Kullback-Leibler
divergence between two pairs of centered Gaussian covariates with
$\langle z_{\alpha i}^{(l-1)}z_{\beta j}^{(l-1)}\rangle=\delta_{ij}g_{l}C^{(l-1)}$
and $\langle z_{\alpha i}^{(l)}z_{\beta j}^{(l)}\rangle=\delta_{ij}\big(C^{(l)}-g_{b}\big)$,
respectively, namely
\begin{align}
 & \mathrm{KL}(\N(0,C^{(l)}-g_{b})||\N(0,g_{l}C^{(l-1)}))\label{eq:KL_divergence}\\
 & =-\big\langle\ln\N(z^{(l)}|0,g_{l}C^{(l-1)}\big)\big\rangle_{z^{(l)}\sim\N(0,C^{(l)}-g_{b})}+\big\langle\ln\N(z^{(l)}|0,C^{(l)}-g_{b}\big)\big\rangle_{z^{(a)}\sim\N(0,C^{(a)}-g_{b})}\nonumber \\
 & =\frac{N}{2g_{l}}\,\tr\,[C^{(l-1)}]^{\T}(C^{(l)}-g_{b})+\frac{N}{2}\,\ln\det\big(C^{(l-1)}\big)-\frac{N}{2}\ln\det\big(C^{(l)}-g_{b}\big)+\mathrm{const.}\,,\nonumber 
\end{align}
where the factors $N$ result from the $z_{\alpha i}$ being i.i.d.
in $i=1,\ldots,N$. Apart from constant terms that we dropped, this
is the same form as \prettyref{eq:gamma_lin_general}. In the case
$g_{b}=0$, the action \prettyref{eq:action_linear} thus reduces
to the main result by \cite{Yang23_39380}, their Eq. (1), when setting
all relative layer width $\nu_{\ell}=1$ as assumed here and using
$K=\mathrm{id}$, valid for deep kernel machines. Our approach is
thus consistent with theirs, if we study deep linear networks. Note
that our general result, the action \prettyref{eq:action_C}, is valid
for deep non-linear networks.

As for the non-linear networks considered in the main text, we may
derive the pair of forward-backward equations for the kernel adaptation
in the linear case. The forward iteration \prettyref{eq:forward_linear}
can be rewritten as
\begin{align}
C^{(l)} & =g_{b}+g_{l}\,C^{(l-1)}\,\big(\I-2\frac{g_{l}}{N}\,\tC^{(l)}C^{(l-1)}\big)^{-1}.\label{eq:forward_linear_final}
\end{align}
The backward equation for $\tC$ arises from computing the MAP estimate
of $C$ from \prettyref{eq:action_linear} as $\partial\S(C)/\partial C_{\alpha\beta}^{(l)}\stackrel{!}{=}0$,
which for $l=L$ yields \prettyref{eq:C_tilde_final} and for $1\le l<L$
evaluates to (cf. \prettyref{eq:stat_intermediate})
\begin{align*}
0 & \stackrel{!}{=}\frac{\partial}{\partial C_{\alpha\beta}^{(l)}}\,\big(\Gamma(C^{(l)}|C^{(l-1)})+\Gamma(C^{(l+1)}|C^{(l)})\big)\\
 & =\tC_{\alpha\beta}^{(l)}-\frac{\partial}{C_{\alpha\beta}^{(l)}}\W(\tC^{(l+1)}|C^{(l)}),
\end{align*}
with the explicit form \prettyref{eq:W_linear} written as $\W(\tC^{(l+1)}|C^{(l)})=\sum_{\alpha\beta}\tilde{C}_{\alpha\beta}^{(l+1)}g_{b}-\frac{N}{2}\,\ln\det\,\big(\I-2\frac{g_{l+1}}{N}\,\tC^{(l+1)}C^{(l)}\big)$
so that
\begin{align}
\tC^{(l)} & =g_{l+1}\,\tC^{(l+1)}\,\big(\I-2\frac{g_{l+1}}{N}\,\tC^{(l+1)}C^{(l)}\big)^{-1}.\label{eq:backward_linear_final}
\end{align}
Note that the form of the forward equation \prettyref{eq:forward_linear_final}
and the backward equation \prettyref{eq:backward_linear_final} show
a symmetry such that $[g_{l}\tC^{(l)}]^{-1}\tC^{(l-1)}=[g_{l}C^{(l-1)}]^{-1}\,(C^{(l)}-g_{b})=\big(\I-2\frac{g_{l}}{N}\,\tC^{(l)}C^{(l-1)}\big)^{-1}$.

\begin{figure*}[t]
\includegraphics{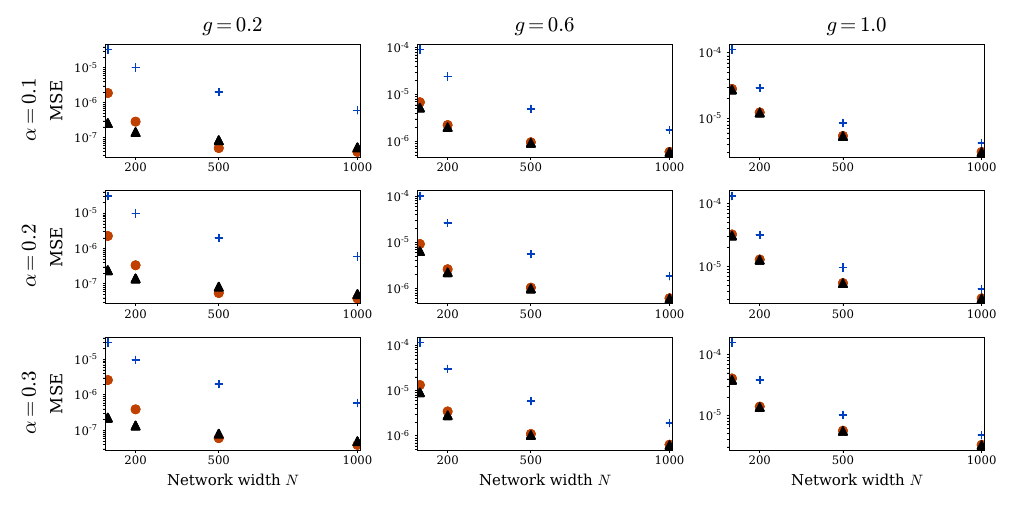}\vspace{0.2in}

\centering{}\caption{Dependence on the ratio between training samples and network width
$\alpha=P/N$ for linear single-hidden layer network. The mean squared
error difference $\mathrm{MSE}(C,C_{\mathrm{emp}})=1/D^{2}\sum_{\alpha,\beta=1}^{D}(C_{\alpha\beta}-C_{\alpha\beta}^{\mathrm{emp}})^{2}$
measures kernel adaptation relative to the numerically sampled kernels.
The $\mathrm{MSE}$ of the data-dependent kernels (blue: NNGP; red:
linear approximation; black: full theory) shows that the feature-corrected
kernels are consistently closer to the empirical kernel than the NNGP
kernel. Parameters: Ising task $\Delta p=0.2,\,D=1000,\,L=1,\,g_{l}=\{0.2,\,0.6,\,1.0\},\,\kappa=0.001,\,\sigma_{b}^{2}=0.05$
for $N=100,\,200,\,500,\,1000$. Results are averaged over $10$ training
data sets and error bars indicate standard deviation.\label{fig:alpha_scan}}
\vspace{0.2in}
\end{figure*}

To test the behavior of linear networks, we use a linearly separable
Ising task: Each pattern $x_{\alpha}$ in the Ising task is $D$-dimensional
and $x_{\alpha i}\in\{\pm1\}$. If the pattern belongs to class $-1$,
each $x_{\alpha i}$ realizes $x_{\alpha i}=+1$ with a probability
of $p_{1}=0.5-\Delta p$ and the value $x_{\alpha i}=-1$ with $p_{2}=0.5+\Delta p$.
The value for each pattern element $x_{\alpha i}$ is drawn independently.
If the pattern belongs to class $+1$, the probabilities for $x_{\alpha i}=1$
and $x_{\alpha i}=-1$ are inverted. The task separability increases
with larger $\Delta p$. In \prettyref{fig:alpha_scan} we plot the
mean squared error difference between the numerically sampled kernels
and the feature-corrected kernels from theory, the NNGP kernels and
the linear approximation in $\tilde{C}$ of the feature corrections
for a linear single-hidden layer network as a function of the ratio
$\alpha=P/N$ between the number of training samples $P$ and network
width $N$. The feature-corrected kernels from the full theory converge
to the empirically measured kernels when increasing the network width
$N$ while keeping $\alpha=P/N$ fixed, showing that the large deviation
result becomes more and more precise. The linear approximation in
$\tC$ yields only a slightly higher error, justifying this approximation.
The deviation between the NNGP and the empirical kernel is consistently
higher, showing that feature corrections remain important in the proportional
limit.

\section{Adaptation towards the target in linear networks\label{app:target_adaptation}}

To gain more intuition into the adaptation of the kernels towards
the target, we may investigate a linear network. To this end consider
the second-order cumulant expansion of $\W$ given by \eqref{eq:W_linear}
as
\begin{align}
\W(\tC^{(l)}|C^{(l-1)}) & =\sum_{\alpha\beta}\tilde{C}_{\alpha\beta}^{(l)}g_{b}+g_{l}\,\tC_{\alpha\beta}^{(l)}C_{\alpha\beta}^{(l-1)}+\frac{g_{l}^{2}}{2N}\,\tC_{\alpha\beta}^{(l)}\,\big(C_{\alpha\gamma}^{(l-1)}C_{\beta\delta}^{(l-1)}+C_{\alpha\delta}^{(l-1)}C_{\beta\gamma}^{(l-1)}\big)\,\tC_{\gamma\delta}^{(l)}+\order(\tC^{3}),\label{eq:W_lin_expanded}
\end{align}
where summation over repeated indices on the right are implied and
where we used that $\langle h_{\alpha}h_{\beta},h_{\gamma}h_{\delta}\rangle^{c}=\langle h_{\alpha}h_{\beta}h_{\gamma}h_{\delta}\rangle-\langle h_{\alpha}h_{\beta}\rangle\,\langle h_{\gamma}h_{\delta}\rangle$,
decomposed the fourth moment into cumulants by Wick's theorem and
what remains are the pairings $\langle h_{\alpha}h_{\gamma}\rangle\,\langle h_{\beta}h_{\delta}\rangle+\langle h_{\alpha}h_{\delta}\rangle\,\langle h_{\beta}h_{\gamma}\rangle=C_{\alpha\gamma}\,C_{\beta\delta}+C_{\alpha\delta}\,C_{\beta\gamma}$.
The stationarity condition \eqref{eq:forward_linear} with \eqref{eq:W_lin_expanded}
takes the form
\begin{align*}
C_{\alpha\beta}^{(l+1)} & =g_{b}+g_{l+1}\,C_{\alpha\beta}^{(l)}+2\,\frac{g_{l+1}^{2}}{N}\,C_{\alpha\gamma}^{(l)}\,\tC_{\gamma\delta}^{(l+1)}\,C_{\delta\beta}^{(l)}.
\end{align*}
Considering the case $g_{b}=0$ in the following, the correction term
comes with a factor $N^{-1}$, so we may approximate within the correction
term $g_{l+1}C^{(l)}\simeq C^{(l+1)}+\order(N^{-1})$. The correction
term in the last layer $l=L$ in this approximation becomes
\begin{align*}
C_{\alpha\beta}^{(L)} & \simeq g_{L}\,C_{\alpha\beta}^{(L-1)}+\frac{2}{N}\,C_{\alpha\gamma}^{(L)}\,\tC_{\gamma\delta}^{(L)}C_{\delta\beta}^{(L)}+\order(N^{-1}).
\end{align*}
Now assuming $\kappa=0$ (no regularization, training without noise)
and inserting the form of $\tC^{(L)}=\frac{1}{2}(C^{(L)})^{-1}YY^{\T}(C^{(L)})^{-1}-\frac{1}{2}(C^{(L)})^{-1}$
given by \eqref{eq:C_tilde_final}, we obtain the correction term
\begin{align*}
\frac{2}{N}\,C_{\alpha\gamma}^{(L)}\,\tC_{\gamma\delta}^{(L)}C_{\delta\beta}^{(L)} & =\frac{2}{N}\,C_{\alpha\gamma}^{(L)}\,\big[\frac{1}{2}(C^{(L)})^{-1}YY^{\T}(C^{(L)})^{-1}-\frac{1}{2}(C^{(L)})^{-1}\big]_{\gamma\delta}\,C_{\delta\beta}^{(L)}\\
 & =\frac{1}{N}\big(YY^{\T}-C^{(L)}\big),
\end{align*}
so that we get the final result
\begin{align}
C_{\alpha\beta}^{(L)} & \simeq g_{L}\,C_{\alpha\beta}^{(L-1)}+\frac{1}{N}\big(YY^{\T}-C^{(L)}\big)_{\alpha\beta}.\label{eq:final_kernel_adapation_YY}
\end{align}
This shows that correction term tends to push the kernel into the
rank-one direction of the target $YY^{\T}$ in order to increase the
log-likelihood of the data.

\section{Relation to the Neural Tangent Kernel\label{app:ntk}}

This section links our work to the neural tangent kernel (NTK). This
material only serves the purpose to recast the known results on the
NTK \cite{Jacot18_8580,Lee19_neurips} into our setting and notation.
We here consider the specific case of the NTK for the squared error
loss function
\begin{align}
\mathcal{\cL}(f;Y) & =\frac{1}{2}\sum_{\alpha=1}^{P}\|f_{\alpha}-y_{\alpha}\|^{2}\label{eq:SQE_loss}
\end{align}
and the network architecture is given by
\begin{align}
h_{\alpha}^{(0)} & =W^{(0)}x_{\alpha},\nonumber \\
h_{\alpha}^{(l)} & =W^{(l)}\phi\left(h_{\alpha}^{(l-1)}\right),\quad l=1,\ldots,L,\label{eq:SecTheory_NetworkArchitecture_NTK}\\
f_{\alpha} & =h_{\alpha}^{(L)}\in\bR,\nonumber 
\end{align}
where to simplify notation we set the biases to zero -- an extension
to non-zero biases is of course possible. To further simplify notation
and to connect the calculations to the main results of the current
work, we assume that the widths for all layers are the same and are
denoted as $N$. In the NTK architecture the weights are scaled as
$W^{(l)}=w^{(l)}/\sqrt{N}$ with $\order(1)\sim w^{(l)}\sim\N(0,g_{l})$
at initialization. For simplicity, we also assume the same dimension
for the data $x_{\alpha}\in\bR^{N}$ here. The $w^{(l)}$ are considered
the trainable parameters which implies that the gradient is multiplied
by $1/\sqrt{N}$ and is hence reduced for large networks. This leads
to the weights $w^{(l)}$ not departing strongly from their initialization.
For equal layer widths the gradient scaling with $1/\sqrt{N}$ corresponds
to using a rescaled loss function $\bar{\mathcal{\cL}}=\mathcal{\cL}/\sqrt{N}$,
so the NTK studies the learning dynamics
\begin{align}
\partial_{t}W & =-\nabla_{W}\,\mathcal{\bar{\cL}}.\label{eq:NTK_learning}
\end{align}
In the following we will only make the link to the NTK right after
initialization, because it has been shown that on the limit $N\to\infty$
the NTK stays constant over training \cite{Jacot18_8580}. As shown
in \cite{Lee19_neurips} the NTK at initialization is equivalent to
linearizing the network outputs $f_{\alpha}$ around a set of initial
weights. This assumption corresponds to assuming that the trained
weights do not depart strongly from their initial point. We will show
here that the framework of Bayesian inference we employ here reduces
to the NTK at initialization under the additional assumption of such
a linear dependence of the network output on the parameters of the
network. This corresponds replacing the mapping between inputs $X\in\bR^{(P+1)\times N}$
and outputs $f(X|W)\in\bR^{P+1}$ implied by \eqref{eq:SecTheory_NetworkArchitecture_NTK}
by a linearized dependency on the parameters $W=\{w_{ij}^{(0)}....w_{ij}^{(L)}\}$
\begin{align}
f(X|W) & \simeq f(X|W_{0})+\nabla f(X|W_{0})\,\omega\label{eq:linearization_in_weights}\\
 & =:f_{0}+\nabla f\,\omega,\nonumber \\
\omega & =W-W_{0},\nonumber 
\end{align}
where $\bR^{(P+1)}\ni\nabla f(X|W_{0})\,\omega:=\sum_{l,ij}\frac{\partial f(X|W)}{\partial w_{ij}^{(l)}}\big|_{W_{0}}\,\omega_{ij}^{(l)}$
and the $\omega$ denote the deviations of the weights from their
initial values. Here $X\in\bR^{(P+1)\times N}$ is the matrix of all
$1\le\alpha\le P+1$ data points, corresponding to $P$ training points
and one test point $\alpha=\ast$ and $\nabla f\in\bR^{(P+1)\times L\,N^{2}+N}$
is the Jacobi matrix of the corresponding network outputs with regard
to the $L\,N^{2}+N$ weight parameters of the network \eqref{eq:SecTheory_NetworkArchitecture_NTK}.

In contrast to \eqref{eq:NTK_learning}, the main part of our work
considers training with a stochastic learning dynamics with weight
decay (see \prettyref{app:langevin})
\begin{align}
\partial_{t}W(t) & =-\gamma W(t)-\nabla\cL(f(X,W(t));Y)+\sqrt{2T}\zeta(t)\,,\label{eq:Appendix_SingleParameter_LSGD_NTK}\\
\big\langle\zeta_{i}(t)\zeta_{j}(s)\big\rangle & =\delta_{ij}\delta(t-s).\nonumber 
\end{align}
We recover NTK training \eqref{eq:NTK_learning} by first changing
the time scale by a factor $\tau=\sqrt{N}$ as
\begin{align}
\tau\,\partial_{t}W(t) & =-\gamma W(t)-\nabla\cL(f(X,W(t));Y)+\sqrt{2T\tau}\,\zeta(t)\,,\label{eq:Appendix_SingleParameter_LSGD_NTK_2}\\
\big\langle\zeta_{i}(t)\zeta_{j}(s)\big\rangle & =\delta_{ij}\delta(t-s).\nonumber 
\end{align}
Since the stationary distribution is invariant under a change of time
scale, both dynamics \eqref{eq:Appendix_SingleParameter_LSGD_NTK}
and \eqref{eq:Appendix_SingleParameter_LSGD_NTK_2} obey the same
stationary distribution. Ultimately we need to set the temperature
$T$ and the weight decay $\gamma$ to zero. We will take $\gamma=0$
immediately, but leave $T$ finite for the intermediate steps of the
calculation and only take the limit in the end. In addition we will
use the linearization \eqref{eq:linearization_in_weights} and therefore
study the dynamics of the $\omega(t)$ following from \eqref{eq:Appendix_SingleParameter_LSGD_NTK_2}
by dividing by $\tau$ as
\begin{align*}
\partial_{t}\omega(t) & =-\nabla_{\omega(t)}\,\mathcal{\bar{L}}(f_{0}+\nabla f\,\omega(t)\,;\,Y)+\sqrt{2T/\tau}\,\zeta(t).
\end{align*}
The stationary distribution of $\omega(t)$ under this dynamics obeys
\begin{align*}
p_{0}(\omega|W_{0}) & \propto\exp\big(-\frac{\tau}{T}\,\bar{\mathcal{L}}(f_{0}+\nabla f\,\omega;\,Y)\big),
\end{align*}
which for the quadratic loss function \eqref{eq:SQE_loss} is a Gaussian
distribution in $\omega$. The resulting joint distribution of network
outputs $f$ and labels $Y$ is then Gaussian, too, because of the
affine linear dependence of $f$ on $\omega$ in \eqref{eq:linearization_in_weights}.
It corresponds to the network prior we compute in the main text \eqref{eq:network_prior}
and here takes the form
\begin{align*}
p(Y,f|X,W_{0})\propto & \int d\omega\,\exp\big(-\frac{\tau}{T}\,\bar{\mathcal{L}}(f;Y)\big)\,\delta(f-f_{0}-\nabla f\,\omega))\\
= & \int d\omega\,\exp\big(-\frac{1}{T}\,\cL(f;Y)\big)\,\delta(f-f_{0}-\nabla f\,\omega)).
\end{align*}
To investigate the statistics of the output conditioned on the training
labels $Y$, it is easiest to introduce the cumulant-generating function
for the conditional $p(f|Y,X,W_{0}):=p(Y,f|X,W_{0})/\int df\,p(Y,f|X,W_{0})$
with $j^{\T}f=\sum_{\alpha=1}^{P+1}j_{\alpha}f_{\alpha}$
\begin{align}
\W(j|Y,X,W_{0}) & =\ln\,\frac{\int df\,p(Y,f|X,W_{0})\,e^{j^{\T}f}}{\int df\,p(Y,f|X,W_{0})}\label{eq:cum_gen_NTK}\\
 & =\ln\,\big\langle e^{j^{\T}f}\big\rangle_{f\sim p(Y,f|X,W_{0})}+\const\nonumber \\
 & =\ln\,\int df\int d\omega\,\exp\big(j^{\T}f-\frac{1}{2T}\,\|Y-f\|_{P}^{2}\big)\,\delta\big(f-f_{0}-\nabla f\,\omega\big)+\const\nonumber \\
 & =\ln\,\int d\omega\,\exp\big(j^{\T}(f_{0}+\nabla f\,\omega)-\frac{1}{2T}\,\|Y-f_{0}-\nabla f\,\omega\|_{P}^{2}\big)+\const\nonumber \\
 & =\ln\,\int d\omega\,\exp\big(j^{\T}(f_{0}+\nabla f\,\omega)-\frac{1}{2T}\omega^{\T}\,\nabla f^{\T}\nabla f\,\omega+\frac{1}{T}\,(Y-f_{0})^{\T}\,\nabla f\,\omega\big)+\const\nonumber \\
 & =\ln\,\int d\omega\,\exp\big(j^{\T}f_{0}+\big(j^{\T}\nabla f+\frac{1}{T}\,(Y-f_{0})^{\T}\nabla f\big)\,\omega-\frac{1}{2T}\omega^{\T}\,\nabla f^{\T}\nabla f\,\omega\big)+\const\nonumber \\
 & =j^{\T}f_{0}+\frac{T}{2}\,\big(j^{\T}\nabla f+\frac{1}{T}\,(Y-f_{0})^{\T}\,\nabla f\big)\,\big[\nabla f^{\T}\nabla f\big]^{-1}\big(j^{\T}\nabla f+\frac{1}{T}\,(Y-f_{0})\,\nabla f\big)^{\T}+\const,\nonumber 
\end{align}
where we performed the Gaussian integral over the $\omega\in\bR^{(L\,N^{2}+N)}$
in the last step and dropped all constant terms (independent of $j$)
along the way. Note that here the norm $\|Y-f\|_{P}^{2}$ is with
regard to the $P$ training points only, likewise all inner products
following from it; the only inner products that involve the test point
are those in $j^{\T}\nabla f$ and $j^{\T}f_{0}$. From the latter
form we can read off that the statistics of the output is Gaussian,
because we obtain a polynomial of degree two in $j$; the mean for
the test point $\alpha=\ast$ is hence its linear coefficient
\begin{align}
\mu_{\ast} & =\frac{\partial}{\partial j_{\ast}}W\big|_{j=0}=f_{0,\ast}+\nabla f_{\ast}\,\big[\nabla f^{\T}\nabla f\big]^{-1}\,\nabla f^{\T}(Y-f_{0}),\label{eq:mean_predictor_NTK}
\end{align}
which is in particular independent of $T$, so that the limit $T\to0$
exists. The variance is given by the term $\propto j^{2}$ which is
linear in $T$ and hence vanishes for $T\to0$. To recover the NTK
result in the known form, we may use that for any matrix $X$ associativity
holds, so that
\begin{align*}
(X^{\T}\,X)\,X^{\T} & =X^{\T}\,(X\,X^{\T}),
\end{align*}
from which follows by multiplying with $(X^{\T}X)^{-1}$ from left
and by $(XX^{\T})^{-1}$ from right
\begin{align*}
X^{\T}\,(X\,X^{\T})^{-1} & =(X^{\T}\,X)^{-1}\,X^{\T}.
\end{align*}
We may use this to rewrite the mean of the predictor in \eqref{eq:mean_predictor_NTK}
as
\begin{align}
\mu_{\ast}=\frac{\partial}{\partial j_{\ast}}W\big|_{j=0} & =f_{0,\ast}+\nabla f_{\ast}\nabla f^{\T}\,\big[\nabla f\nabla f^{\T}\big]^{-1}\,(Y-f_{0}),\label{eq:NTK_predictor}
\end{align}
where the NTK kernel
\begin{align}
\Theta_{\alpha\beta} & =[\nabla f\nabla f^{\T}]_{\alpha\beta}\equiv\sum_{l,ij}\,\frac{\partial f_{\alpha}}{\partial W_{ij}^{(l)}}\,\frac{\partial f_{\beta}}{\partial W_{ij}^{(l)}}\label{eq:NTK_kernel}
\end{align}
for $1\le\alpha,\beta\le P+1$ appears; specifically, the matrix $\big[\nabla f\nabla f^{\T}\big]_{1\le\alpha,\beta\le P}$
is with regard to the $P$ training points only (inherited from the
norm $\|\ldots\|_{P}^{2}$), while $[\nabla f_{\ast}\nabla f^{\T}]_{1\le\beta\le P}$
is a vector of dimension $P$, fixing the left index to the training
point $\alpha=\ast$ (coming from the derivative by $j_{\ast}$).
The expression \eqref{eq:NTK_predictor} is the stationary point of
the NTK predictor for the linearized network (cf. Eqs. (10)-(11) in
\cite{Lee19_neurips}).

The original work (Theorem 2 in Sec 4.2 of \cite{Jacot18_8580}) has
shown that in the limit $N\to\infty$ the NTK stays constant over
training, which implies that the expressions obtained here remain
valid throughout training.

This shows that the posterior we compute in our general framework
is consistent with the NTK if we make the additional assumption that
the dependence of the network output can be linearized with regard
to the trained parameters; such an assumption is justified if the
weights do not depart strongly from their initial values, as in the
NTK setting in the $N\to\infty$ limit. Our general approach does
not require this linearization. In addition, for the NTK we took vanishing
weight decay -- an extension to non-zero weight decay would be possible,
though, still remaining with a Gaussian $\omega$ and hence $f$.
The computation here also shows that we may use a non-zero $T$ in
the training dynamics as we do in the main text and would still obtain
the same result \eqref{eq:NTK_predictor} for the mean of the predictor,
albeit with a non-zero variance that can be read off from \eqref{eq:cum_gen_NTK}.

The conceptually important difference between the NTK kernel $\Theta$
\eqref{eq:NTK_kernel} and the kernels $C$ we study is that the NTK
kernel is agnostic to the training labels $Y$ -- the shape of the
kernel only depends on the network architecture and the data points
$X$, but not on the labels $Y$. Similar to the NNGP the NTK is hence
unable to relate the inputs $X$ and the labels $Y$ and hence does
not capture feature learning. Our work, in contrast, investigates
how kernels are shaped by the joint statistics of $X$ and $Y$ --
this is evident from the fact that the MAP estimate of the kernels
results from an interplay of the likelihood of the labels $\S_{D}$
and the prior term in \eqref{eq:action_C} and is shown by the increase
of the CKA between $C^{(L)}$ and $YY^{\T}$; for linear networks
this increase is shown explicitly in \eqref{eq:final_kernel_adapation_YY}.

\section{Price's theorem\label{app:Generalization-of-Price}}

Consider an expectation value of $f:\mathbb{R}^{N}\rightarrow\mathbb{R}$
over centered jointly Gaussian distributed $x_{i}$ with covariance
$C$
\begin{align*}
\langle f(x)\rangle_{x\sim\N(0,C)}.
\end{align*}
We assume that $f(x)$ grows slower than $e^{x_{i}^{2}}$ for large
$x_{i}$. Rewriting the Gaussian $\N(0,C)$ in terms of its Fourier
transform $\N(0,C)=\Big\{\prod_{j}\int_{-i\infty}^{i\infty}\,\frac{d\tilde{x}_{j}}{2\pi i}\Big\}\exp\big(-x^{\T}\tilde{x}+\frac{1}{2}\tilde{x}^{\T}C\tilde{x}\big)$
one obtains
\begin{align*}
\langle f(x)\rangle_{x\sim\N(0,C)} & =\prod_{j}\Big\{\int_{-\infty}^{\infty}dx_{j}\,\int_{-i\infty}^{i\infty}\,\frac{d\tilde{x}_{j}}{2\pi i}\Big\}\,f(x)\,\exp\big(-x^{\T}\tilde{x}+\frac{1}{2}\tilde{x}^{\T}C\tilde{x}\big),
\end{align*}
which yields the property
\begin{align*}
\frac{\partial}{\partial C_{kl}}\langle f(x)\rangle_{x\sim\N(0,C)} & =\prod_{j}\Big\{\int_{-\infty}^{\infty}dx_{j}\,\int_{-i\infty}^{i\infty}\,\frac{d\tilde{x}_{j}}{2\pi i}\Big\}\,f(x)\,\frac{1}{2}\,\tilde{x}_{k}\tilde{x}_{l}\,\exp\big(-x^{\T}\tilde{x}+\frac{1}{2}\tilde{x}^{\T}C\tilde{x}\big).
\end{align*}
One notices that one may replace both occurrences of $\tilde{x}_{i}\to-\partial/\partial x_{i}$
under the integral: integrating by parts twice and using the assumption
that $f$ grows slower than $e^{x_{i}^{2}}$ for large $x_{i}$ so
that boundary terms vanish, one obtains
\begin{align}
\frac{\partial}{\partial C_{kl}}\langle f(x)\rangle_{x\sim\N(0,C)} & =\prod_{j}\Big\{\int_{-\infty}^{\infty}dx_{j}\,\int_{-i\infty}^{i\infty}\,\frac{d\tilde{x}_{j}}{2\pi i}\Big\}\,\frac{1}{2}\,\Big\{\frac{\partial}{\partial x_{k}}\,\frac{\partial}{\partial x_{l}}\,f(x)\Big\}\,\exp\big(x^{\T}\tilde{x}+\frac{1}{2}\tilde{x}^{\T}C\tilde{x}\big)\nonumber \\
 & =\frac{1}{2}\,\big\langle f_{kl}^{(2)}\big\rangle_{x\sim\N(0,C)},\label{eq:price_general-2}
\end{align}
where $f_{kl}^{(2)}$ is the Hessian of $f$. This expression is known
as Price's theorem \cite{Price58_69,PapoulisProb4th}. Note that sometimes
the theorem is only stated for derivatives by $C_{k\neq l}$ only.

This theorem can be used to rewrite
\begin{equation}
\frac{\partial}{\partial C_{\alpha\beta}^{(l)}}\Big\langle\exp\big(\frac{g_{l+1}}{N}\phi^{(l)\T}\tC^{(l+1)}\phi^{(l)}\big)\Big\rangle_{h^{(l)}\sim\N(0,C^{(l)})}
\end{equation}
to obtain an expression for $\frac{\partial}{\partial C_{\alpha\beta}^{(l)}}\,\W(\tC^{(l+1)}|C^{(l)})$
in \prettyref{eq:CFL_Full_Tilde_Kernel_Propagation} (see \prettyref{app:Maximum-a-posteriori}).

\section{Langevin stochastic gradient descent\label{app:langevin}}

To compare theoretical results to real networks we sample numerically
from the posterior of networks that have been conditioned on the training
data $X=(x_{\alpha})_{\alpha=1,\ldots,P},Y=(y_{\alpha})_{\alpha=1,\ldots,P}$.
We therefore train the network \eqref{eq:SecTheory_NetworkArchitecture}
using Langevin stochastic gradient descent (LSGD). According to \cite{Naveh21_064301}
evolving parameters $\Theta$ with the stochastic differential equation
\begin{align}
\partial_{t}\Theta(t) & =-\gamma\Theta(t)-\nabla_{\Theta}\mathcal{L}(\Theta(t);Y)+\sqrt{2T}\zeta(t),\label{eq:Appendix_SingleParameter_LSGD}\\
\big\langle\zeta_{i}(t)\zeta_{j}(s)\big\rangle & =\delta_{ij}\delta(t-s),\nonumber 
\end{align}
with the squared error loss $\mathcal{L}(\Theta;Y)=\sum_{\alpha=1}^{P}(f_{\alpha}(\Theta)-y_{\alpha})^{2}$
and $f_{\alpha}(\Theta)$ denoting the network output for sample $\alpha$,
leads to sampling from the equilibrium distribution for $\Theta$
for large times $t$ which reads
\begin{equation}
\lim_{t\rightarrow\infty}p\left(\Theta(t)\right)\sim\exp\left(-\frac{\gamma}{2T}\|\Theta\|^{2}-\frac{1}{T}\mathcal{L}(\Theta;Y)\right).
\end{equation}
The equilibrium distribution may be derived from the Fokker-Planck
equation \cite{Risken96} for the density of $\Theta$. Conversely,
this implies a density for the output
\begin{align}
p(Y|X)\propto & \int d\Theta\,\exp\big(-\frac{\gamma}{2T}\,\|\Theta\|^{2}-\frac{1}{T}\,\|f-Y\|^{2}\big)\\
\propto & \Big\langle\exp\big(-\frac{1}{T}\,\|f-Y\|^{2}\big)\Big\rangle_{\Theta_{k}\stackrel{\text{i.i.d.}}{\sim}\N(0,T/\gamma)}\nonumber \\
\propto & \N(Y|f,T/2)\,\langle\delta\big[f-f(\Theta)\big]\rangle_{\Theta_{k}\stackrel{\text{i.i.d.}}{\sim}\N(0,T/\gamma)},\nonumber 
\end{align}
which, with $p(f|X)\equiv\langle\delta\big[f-f(\Theta)\big]\rangle_{\Theta_{k}\stackrel{\text{i.i.d.}}{\sim}\N(0,T/\gamma)}$,
is identical to \eqref{eq:network_prior} if one identifies $\kappa=T/2$
with the regularization noise and $T/\ensuremath{\gamma}=g/N$ with
the variance of the parameter $\Theta_{k}$. To implement different
variances in practice, one requires a different weight decay $\gamma$
for each parameter.

The time discrete version of \prettyref{eq:Appendix_SingleParameter_LSGD}
is implemented as
\begin{align}
\Theta_{t} & =\Theta_{t-1}-\eta\left(\gamma\Theta_{t-1}+\nabla_{\Theta}\mathcal{L}(\Theta_{t-1};Y)\right)+\sqrt{2T\eta}\,\zeta_{t},\\
\langle\zeta_{t}\zeta_{s}\rangle & =\delta_{ts},\nonumber 
\end{align}
with standard normal $\zeta_{t}$ and finite time step $\eta$, which
can also be interpreted as a learning rate. To accurately reflect
the time evolution according to \prettyref{eq:Appendix_SingleParameter_LSGD}
the learning rate needs to be small. Hence Langevin stochastic gradient
descent corresponds to full-batch gradient descent with the addition
of i.i.d. distributed standard normal noise and weight decay regularization
\cite{Krogh91}. The value for $\kappa$, which appears in the main
text as the regularizer on the diagonal of the output kernel $C^{(L)}$,
quantifies the tradeoff between the influence of the prior and the
influence of the training data via the loss term. Choosing large $\kappa$
corresponds to large $T=2\kappa$ and hence a large noise in the LSGD,
putting more emphasis on the parameter priors. In contrast, small
regularization values $\kappa$ favor the training data in the loss
in the exponent. When using LSGD to sample from the equilibrium distribution,
it needs to be ensured that the distribution is equilibrated and subsequent
network samples drawn from the distribution are uncorrelated. For
empirical results, we therefore sample networks after an initial warmup
of 50.000 training steps in distances of 1.000 time steps.

\section{Additional details of theory implementation\label{app:details_implementation}}

\subsection{Setting weight variance of input layer}

The response functions $\chi^{l,\rightarrow}$ describe the effect
of a perturbation in the input kernel the kernel in layer $l$. It
depends on the network kernels $C_{\alpha\alpha}^{(k)}$ of all layers
before layer $k$. For simplicity, we set the weight variance of the
input layer $g_{0}$ such that the diagonal elements of the network
kernels $C_{\alpha\alpha}^{(l)}$ are already at their fixed point
value for large depth \cite{Schoenholz17_iclr}. Consequently, the
convergence of the diagonal kernel elements does not influence the
response functions and there remains only one relevant relaxation
scale for the latter.

\subsection{Gaussian integrals}

We solve the self-consistency equations in \prettyref{eq:CFL_Full_Tilde_Kernel_Propagation}
iteratively. This requires computing two-point and four-point Gaussian
integrals. For $\phi=\text{erf}$, we obtain the following analytical
expressions for the two-point integrals
\begin{align*}
\left\langle \phi(h_{\alpha})\phi(h_{\beta})\right\rangle _{h\sim\mathcal{N}(0,C)} & =\begin{cases}
\frac{4}{\pi}\arctan\left(\sqrt{1+4C_{\alpha\alpha}}\right)-1 & \alpha=\beta,\\
\frac{2}{\pi}\arcsin\left(\frac{2C_{\alpha\beta}}{\sqrt{1+2C_{\alpha\alpha}}\sqrt{1+2C_{\beta\beta}}}\right) & \text{else,}
\end{cases}\\
\left\langle \phi^{\prime}(h_{\alpha})\phi^{\prime}(h_{\beta})\right\rangle _{h\sim\mathcal{N}(0,C)} & =\begin{cases}
\frac{4}{\pi}\frac{1}{\sqrt{4C_{\alpha\alpha}+1}} & \alpha=\beta,\\
\frac{4}{\pi}\left(2\left(C_{\alpha\alpha}+C_{\beta\beta}\right)+1+4\left(C_{\alpha\alpha}C_{\beta\beta}-C_{\alpha\beta}^{2}\right)\right)^{-0.5} & \text{else},
\end{cases}\\
\left\langle \phi(h_{\alpha})\phi^{\prime\prime}(h_{\beta})\right\rangle _{h\sim\mathcal{N}(0,C)} & =\begin{cases}
-\frac{8}{\pi}\frac{C_{\alpha\alpha}}{(2C_{\alpha\alpha}+1)\sqrt{4C_{\alpha\alpha}+1}} & \alpha=\beta,\\
-\frac{8}{\pi}\frac{C_{\beta\alpha}}{(2C_{\alpha\alpha}+1)\sqrt{2\left(C_{\alpha\alpha}+C_{\beta\beta}\right)+1+4\left(C_{\alpha\alpha}C_{\beta\beta}-C_{\alpha\beta}^{2}\right)}} & \text{else}.
\end{cases}
\end{align*}
We are not aware of an analytical solution for the appearing four-point
integral $\Big\langle\phi(h_{\alpha})\phi(h_{\beta})\phi(h_{\gamma})\phi(h_{\delta})\Big\rangle_{h\sim\mathcal{N}(0,C)}$.
Therefore, we determine this integral numerically using Monte-Carlo
sampling with $n_{MC}=10^{5}$ samples.

\subsection{Annealing in network width}

In \prettyref{sec:theory} we derived self-consistency equations for
the posterior kernels perturbatively up to linear order in the conjugate
kernels $\tilde{C}^{(l)}$ (see \prettyref{eq:CFL_Full_Tilde_Kernel_Propagation}).
We solve these equations iteratively: i) Initialize $C^{(0)}$ by
\eqref{eq:C_0} and set $\tC=0$ initially. ii) Iterate \eqref{eq:perturbative_forward}
forward until $C^{(L)}$; in the first iteration this step still corresponds
to the NNGP. iii) Determine $\tC^{(L)}$ in the final layer from \eqref{eq:C_tilde_final}.
iv) Propagate $\tC$ backward with \eqref{eq:CFL_Full_Tilde_Kernel_Propagation}
(but using the Gaussian measure $\langle\ldots\rangle_{\N(0,C^{(l)})}$
instead of the non-Gaussian measure $\langle\ldots\rangle_{\cP^{(l)}}$
throughout \eqref{eq:CFL_Full_Tilde_Kernel_Propagation}). Then go
back to step ii) with $\tC\neq0$ and iterate until convergence. To
improve the stability of these iterations, we use a damping parameter
$\gamma=0.5$ and replace per iteration $i$ as
\begin{align*}
C^{(l),i} & \mapsto(1-\gamma)C^{(l),i+1}+\gamma C^{(l),i},\\
\tilde{C}^{(l),i} & \mapsto(1-\gamma)\tilde{C}^{(l),i+1}+\gamma\tilde{C}^{(l),i}.
\end{align*}
When solving these equations iteratively, we use the NNGP kernel as
the starting value. For wide networks and fixed training data $P/N\rightarrow0$,
corrections to the NNGP kernel become small and posterior kernels
are well described by including corrections up to linear order. To
obtain posterior kernels for arbitrary network widths, we use that
corrections are small when determining the posterior kernels based
on the posterior kernels of a slightly wider network: We start from
very wide networks and compute corrections to the NNGP kernel. Then
we use these corrected kernels as the starting point for slightly
narrower networks and repeat until a certain network width (see pseudo
code in \ref{alg:width_annealing}).

\begin{algorithm}[tb]
\caption{Width annealing of kernels}
\label{alg:width_annealing}
\begin{algorithmic}
   \STATE {\bfseries Input:} data $X$, labels $Y$, network widths $\{N_i\}_i$
   \STATE Compute NNGP kernel $C^{(l)}_{\text{NNGP}}$ from data $X$.
   \STATE Set start values to NNGP kernel $C^{(l)}_{\text{init}}=C^{(l)}_{\text{NNGP}}$ and $\tilde{C}^{(l)}_{\text{init}}=0$.
   \FOR{$N$ {\bfseries in} $\{N_i\}_i$}
   \STATE Compute corrected kernels $C^{(l)}_{\text{corr}}=f(C^{(l)}_{\text{init}}, \tilde{C}^{(l)}_{\text{init}}, Y, N)$ and conjugate kernels $\tilde{C}^{(l)}_{\text{corr}}=g(C^{(l)}_{\text{init}}, \tilde{C}^{(l)}_{\text{init}}, Y, N)$.
   \STATE Reset start values $C^{(l)}_{\text{init}}=C^{(l)}_{\text{corr}}$ and $\tilde{C}^{(l)}_{\text{init}}=\tilde{C}^{(l)}_{\text{corr}}$.   \ENDFOR
\end{algorithmic}
\end{algorithm}

\begin{figure}[t]
\vspace{0.2in}

\includegraphics{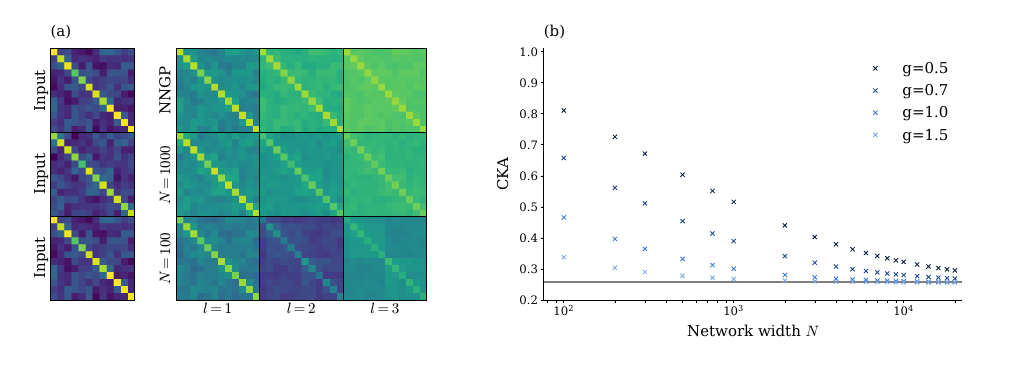}
\centering{}\caption{Annealing in network width to solve self-consistency equations. (a)
Network kernel $C^{(l)}$ across layers $l=1,\,2,\,3$ for different
network width $N$ and $g_{l}=g=0.5$. More narrow networks show stronger
adaptation to the target kernel $YY^{\mathsf{T}}$ across layers.
(b) CKA between network kernels $C^{(L)}$ and target kernel $YY^{\mathsf{T}}$
for different network width. For wide networks, the CKA (blue markers)
remains close to that of the NNGP (solid line). For more narrow networks,
the corrections towards the target kernel and away from the NNGP kernel
increase continuously. The correction strength depends on other network
hyperparameters such as the weight variance $g_{l}$ (increasing from
dark to light). Other parameters: XOR task with $\sigma^{2}=0.4$,
$g_{l}\in\left\{ 0.5,\,0.7,\,1.0,\,1.5\right\} ,\,g_{b}=0.05,\,L=3,\,\kappa=10^{-3},\,P=12$.\label{fig:width_annealing}}
\vspace{0.2in}
\end{figure}

In \prettyref{fig:width_annealing}, we show the CKA between the output
kernel $C^{(L)}$ and target kernel $YY^{\mathsf{T}}$ relative to
the NNGP kernel for annealing in network width.

\section{Centered kernel alignment\label{app:centred_kernel_alignment}}

According to \cite{Canatar22_ieee}, the kernel alignment between
two kernels $A,\,B\in\mathbb{R}^{P\times P}$ is measured by
\[
\frac{\text{Tr}(A\,B)}{\sqrt{\text{Tr}(A\,A)\,\text{Tr}(B\,B)}}.
\]
This corresponds to the cosine similarity between the flattened kernels
and is thus invariant under scaling the kernels by a scalar $A\mapsto aA$.
To remove constant components in the eigendecomposition of the kernel
\cite{Cortes12_jmlr}, we use the centered kernel alignment (CKA):
the kernels are transformed as $A\mapsto HAH$ and $B\mapsto HBH$
with the centering matrix $H=\mathbb{I}-\frac{1}{P}11^{\mathsf{T}}$
where $1$ is the matrix with all ones as elements. Throughout this
work, we study the CKA between network kernels $C^{(l)}$ and the
target kernel $YY^{\mathsf{T}}$.
\end{document}